\documentclass[aps,prx,twocolumn, superscriptaddress, notitlepage, nofootinbib, longbibliography]{revtex4-1}
\usepackage{amsmath,braket,amssymb}
\usepackage[final]{graphicx}
\usepackage{subfigure}
\usepackage[english]{babel}
\usepackage[utf8]{inputenc}
\usepackage{mathtools}
\usepackage{empheq}
\usepackage{tensor} 
\usepackage{cancel}
\usepackage{mathtools}
\usepackage{simplewick}

\usepackage[usenames,dvipsnames]{color}

\usepackage{enumitem}
\usepackage{slashed}
\usepackage{graphicx}
\usepackage{xcolor}

\colorlet{Green}{black!30!green}

\usepackage{tikz}
\usetikzlibrary{calc,fadings,decorations.pathreplacing,shapes,shapes.multipart,arrows,shapes.misc,intersections,positioning,patterns}
\tikzset{arrow data/.style 2 args={%
		decoration={%
			markings,
			mark=at position #1 with \arrow{#2}},
		postaction=decorate}
}
\usepackage{bm}
\usepackage[colorlinks=true,citecolor=blue,linkcolor=blue,urlcolor=blue]{hyperref}
\usepackage{dsfont}
\usepackage{changepage}
\usepackage{array}
\usepackage{soul}
\usepackage[capitalise]{cleveref}
\crefname{section}{Sec.}{Secs.}
\Crefname{section}{Sec.}{Secs.}
\usepackage{xifthen}
\usepackage{xargs}
\usepackage{dynkin-diagrams}
\usepackage{float}
\usepackage[utf8]{inputenc}
\usepackage[normalem]{ulem}

\usepackage{bm}
\renewcommand{\vec}[1]{\boldsymbol{\mathbf{#1}}}

\graphicspath{{./}{./figures/}}

\newcommand{\bit}{\begin{itemize}}
\newcommand{\eit}{\end{itemize}}

\newcommand{\ba}{\begin{align}}
\newcommand{\ea}{\end{align}}
\newcommand{\be}{\begin{equation}}
\newcommand{\ee}{\end{equation}}
\newcommand{\bi}{\begin{itemize}}
\newcommand{\ei}{\end{itemize}}

\newcommand{\dd}{\mathrm{d}}

\newcommand{\id}{\mathbb{I}}
\newcommand{\Tr}{\operatorname{Tr}}

\newcommand{\horror}{\kappa^{\beta,\mathcal{E}}}

\newcommand{\teal}{\color{teal}}

\DeclareMathAlphabet{\mymathbb}{U}{BOONDOX-ds}{m}{n}

\renewcommand{\log}{\ln}

\usepackage{braket}
\usepackage[normalem]{ulem}
\usepackage{comment}

\begin{document}

\title{Designs via Free Probability}

\author{Michele Fava}
\affiliation{Philippe Meyer Institute, Laboratoire de Physique de l’\'{E}cole Normale Sup\'{e}rieure (ENS), Universit\'{e} PSL, 24 rue Lhomond, F-75231 Paris, France}

\author{Jorge  Kurchan}
\affiliation{Laboratoire de Physique de l’\'Ecole Normale Sup\'erieure, ENS, Universit\'e PSL, CNRS, Sorbonne Universit\'e, Universit\'e de Paris, F-75005 Paris, France}

\author{Silvia Pappalardi}
\email{pappalardi@thp.uni-koeln.de}
\affiliation{Institut f\"ur Theoretische Physik, Universit\"at zu K\"oln, Z\"ulpicher Straße 77, 50937 K\"oln, Germany}

\date{\today}

\begin{abstract}
    Unitary Designs have become a vital tool for investigating pseudorandomness since they approximate the statistics of the uniform Haar ensemble. Despite their central role in quantum information,  their relation to quantum chaotic evolution and in particular to the Eigenstate Thermalization Hypothesis (ETH) are still largely debated issues. This work provides a bridge between the latter and $k$-designs through Free Probability theory. 
    First, by introducing the more general notion of  $k$-freeness, we show that it can be used as an alternative probe to designs.
    In turn, free probability theory comes with several tools, useful for instance for the calculation of mixed moments or the so-called $k$-fold quantum channels.
    Our second result is the connection to quantum dynamics. Quantum ergodicity, and correspondingly ETH, apply to a restricted class of physical observables, as already discussed in the literature. In this spirit, we show that unitary evolution with generic Hamiltonians always leads to freeness at sufficiently long times, but only when the operators considered are restricted within the ETH class.
    Our results provide a direct link between unitary designs, quantum chaos and the Eigenstate Thermalization Hypothesis, and shed new light on the universality of late-time quantum dynamics.    
\end{abstract}

\maketitle

\section{Introduction}

The purpose of this paper is to make explicit the connections between three lines of thought: $k$-designs, the `full' version of the Eigenstate Thermalization Hypothesis, and Free Probability.

In recent years, there has been a growing interest in the study of random unitaries and their uniform distribution, the Haar ensemble. 
These concepts are essential in quantum information, with applications ranging from cryptography~\cite{chau2005unconditionally} and quantum key distribution~\cite{renes2004symmetric,hayashi2005reexamination, iblisdir2006optimal}, to randomized benchmarking and measurements~\cite{Emerson_2005, PhysRevLett.108.110503,PhysRevLett.120.050406,PhysRevA.99.052323,huang2020predicting,paini2019approximate,morris2020selective,knips2020multipartite,PhysRevLett.122.120505,Elben_2022}, as well as many others~\cite{low2010pseudorandomness}.
However, sampling unitaries from the Haar ensemble in practical setups is challenging, especially for large Hilbert spaces. To address this issue,  $k$\emph{-designs} have been introduced. These correspond to ensembles of unitaries which approximate the Haar ensemble by reproducing its correlations up a certain order $k$ \cite{delsarte1991spherical, dankert2005efficient, dankert2009exact}.
Random unitaries play a fundamental role also in the study of the chaotic evolution of many-body systems. The reasons are multiple: on one side many universal properties of chaotic many-body systems are understood from random matrix theory, on the other, chaotic evolution is expected to lead the system to uniformly sample Hilbert space at sufficiently long times, as some form of ergodicity. 
Several works in these directions have studied the emergence of design properties in different models, such as random unitary circuits~\cite{Harrow2009,Diniz2011,PhysRevLett.116.170502,Haferkamp2022,PhysRevX.7.021006,Harrow2023,hunterjones2019unitary,brandao2021models,Haferkamp2022a,low2010pseudorandomness,PhysRevA.72.060302,hearth2023unitary} and many-body models with Brownian couplings~\cite{jian2022linear,tiutiakina2023frame}. Other approaches have focused on the random properties of ensembles of states \cite{brandao2021models, pilatowsky2023complete}. In this case, a popular setup is to consider the ensemble as arising from projective measurements on a subsystem,
the so-called deep thermalization \cite{cotler2023emergent, Choi_2023, ho2022exact, claeys2022emergent, ippoliti2022solvable,ippoliti2023dynamical,lucas2022generalized, bhore2023deep, mcginley2022shadow, PRXQuantum.4.030322}.\\

A framework for understanding quantum statistical mechanics is provided by the \emph{Eigenstate Thermalization Hypothesis} (ETH), 
which describes the statistical properties of a class of physical observables $\mathcal A$ when represented in the basis of generic many-body Hamiltonians \cite{deutsch1991quantum, srednicki1999approach}. 
This approach has a well-defined classical limit in Khichin's statistical mechanics \cite{aleksandr1949mathematical}, that, rather than on ergodicity, focuses on observables and their properties in the thermodynamic limit. 
The validity of ETH in local many-body systems has been verified in a plethora of numerical experiments \cite{prosen1999, rigol2008thermalization, biroli2010effect, polkovnikov2011colloquium, ikeda2013finite, steinigeweg2013eigenstate, alba2015eigenstate, beugeling2015off, luitz2016long, luitz2016anomalous, leblond2020eigenstate, brenes2020eigenstate, fritzsch2021eigenstate, garratt2021pairing}, making ETH one of the most established paradigms for thermalization in quantum systems. 

Recently, the introduction of out-of-time order correlators (OTOC) as a probe of quantum chaos and information scrambling \cite{KitaTalk, Maldacena2016bound, hosur2016chaos} has emphasized the importance of higher-order correlations to understand the connection between these two concepts. On the one hand, higher-order OTOCs encoding correlations of order $k$ and 
naturally raised the question of whether evolution may give rise to $k$-designs,
as discussed by Roberts and Yoshida in Ref.\cite{roberts2017chaos}. 
On the other hand, the non-trivial behaviour of OTOC motivated the development of a ``full'' version of ETH \cite{foini2019eigenstate} which encompasses all correlations between matrix elements of observables.  
This perspective has attracted a lot of interest from the many-body~\cite{foini2019eigenstate2, chan2019eigenstate, murthy2019bounds, richter2020eigenstate, wang2021eigenstate, brenes2021out,  dymarsky2022bound, nussinov2022exact} to the high-energy communities \cite{sonner2017eigenstate, jafferis2022matrix,jafferis2022jt}.
More recently, a direct connection between the full ETH and multi-time correlation functions has been achieved using the language of \emph{Free Probability}~\cite{pappalardi2022eigenstate}. 
This is a generalization of classical probability theory to non-commuting variables,  where ``independence'' between random variables becomes ``free independence'' -- or ``freeness''--, standard cumulants become ``free cumulants'' and many other concepts are revised to be applied to non-commuting objects \cite{voiculescu1991limit, speicher1997free, speicher2009free}. 
When dealing with large $D\times D$ (random) matrices, \emph{asymptotic} freeness applies, valid when the size of the matrix $D$ diverges.\\
 In the language of Free Probability, the full eigenstate-thermalization hypothesis  takes a very simple form since the statistics of matrix elements  
is directly encoded in terms of the thermal free cumulants~\cite{pappalardi2022eigenstate} ---  a property that  was later confirmed numerically~\cite{Pappalardi2023quantum}.

Interestingly, free probability was initially anticipated in planar field theory \cite{brezin1978planar, cvitanovic1981planar, cvitanovic1982planar} and it has afterward emerged in several other branches of quantum mechanics both at a technical and a conceptual level \cite{ebrahimi2016combinatorics, collins2016random, bellitti2019hamiltonian, shapourian2021entanglement, wang2023beyond}. Lately, it has also appeared in the description of mesoscopic models of quantum transport \cite{Bauer_2017, Hruza_2023, bauer2023bernoulli, bernard2023exact} and for studying the thermalization of Wigner matrices \cite{cipolloni2022thermalisation}.
All these findings suggest that Free Probability, which focuses on non-commuting objects, could have a wider range of applicability in the study of many-body quantum systems.\\

In this work, we use Free Probability to study unitary $k$-designs and to elucidate the connections to the long-time evolution of large chaotic quantum systems, as summarized in Fig.\ref{fig_summary}. Our starting point is the definition of \emph{asymptotic $k$-freeness} which relaxes the one of freeness, restricting it to order $k$. 
Building on this, we show that $k$-designs imply asymptotic $k$-freeness for every set of operators, in the limit of large matrix size.
As a consequence, our result provides many useful tools for the study of designs in quantum information. These include an asymptotic expansion for quantum channels in terms of free cumulants or a closed expression to compute OTOC.  \\
Secondly, we consider Hamiltonian time evolution and we restrict ourselves to physical ETH observables $\mathcal A$ based on local operators, and in particular, those having a good semiclassical limit. A relation between time-evolution and designs restricted to $\mathcal A$ was already discussed under the name of ``partial unitary design'' by Kaneko, Iyoda and Sagawa and in Ref.\cite{kaneko2020characterizing}. 
In the same ETH spirit, we employ a Free Probability approach which shifts the attention from designs to the relation between different observables. Importantly, we shall see that the set of unitaries obtained by time evolution implies  
arbitrary asymptotic $k$-freeness, restricted to $\mathcal A$. This result holds at any finite temperature and is mathematically rigorous upon infinite time average. 
In physical chaotic many-body Hamiltonians, we shall argue that $k$-freeness is obtained at a finite \emph{free}-$k$ time $t_k$ which depends on the Hamiltonian microscopic and on the observables.


The results of this work provide a novel perspective on the relationship between unitary design and the long-time universality of chaotic evolution. Our findings show that at sufficiently long times Hamiltonian dynamics are as good as designs in making physical observables free.

The rest of the paper is organized as follows. In Sec.\ref{sec_summary} we present an overview of the content of the paper, introducing the concepts, summarizing our new results, with a comment on the differences with the other works in the literature. To make the paper self-contained, in Sec.\ref{sec_notion} we provide a pedagogical overview of the notions and notations that are studied in this paper. The next two sections contain the bulk of new results: in Sec.\ref{sec_result_des}, we discuss $k$-designs via Free Probability and in Sec.\ref{sec_result_eth} we provide the bridge between ETH and long-time dynamics. Finally, Sec.\ref{sec_conclusion} is devoted to the conclusions and the open questions. 

\tableofcontents

\section{Summary}
\label{sec_summary}
\begin{figure}[t]
\hspace{-.4cm}
\includegraphics[width=1.1\columnwidth]{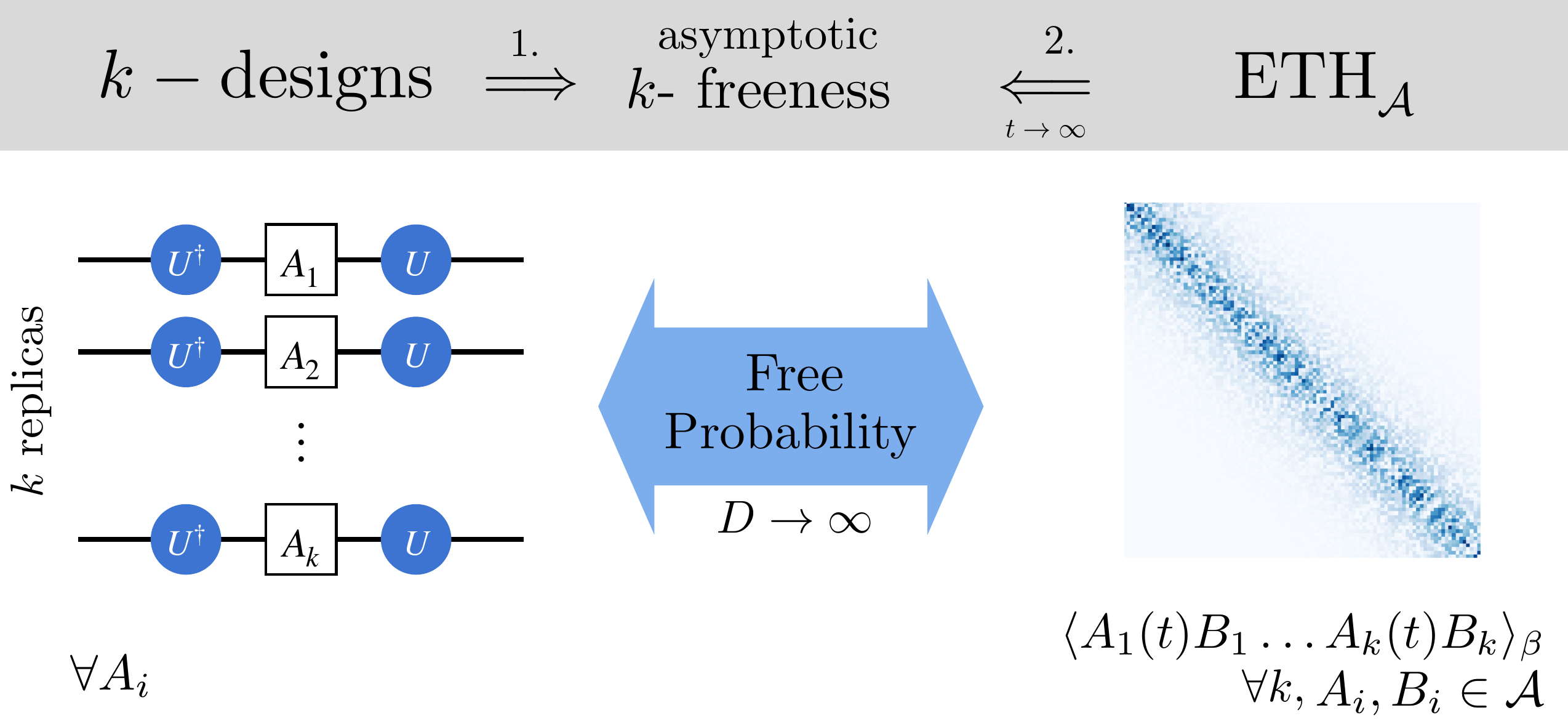}%
\caption{Bridge between $k$-designs (on the left) and the Eigenstate Thermalization Hypothesis (on the right) provided by Free Probability. The connection steps on the definition of asymptotic $k$-freeness valid in the limit of large Hilbert space dimension $\text{dim}\mathcal H=D\to \infty$.  As a first result, we show that $k$-designs imply $k$-freeness \text{asymptotically}, i.e. in the limit of large $D$, for every observable. Secondly, we show that generic unitary time evolution implies $k$-freeness  at sufficiently large times, but only for observables restricted to the subclass $\mathcal A$ of physical observables for which ETH applies.}
\label{fig_summary}
\end{figure}

In this section, we aim to give the reader a summary of the content and results presented in this paper.  Our main
findings are summarized and illustrated in Fig.\ref{fig_summary} and discussed below.

\subsection*{Full Eigenstate Thermalization Hypothesis}
In recent years there has been a renewed interest in quantum chaos, spurred by new ideas coming
from the black-hole/holography community on the one side, and from Quantum Information on the other. Unlike the approaches in the 1970's and 80's, which were mostly concentrated
on semiclassical tools, the interest now is directed on the thermodynamic limit.

In a very influential paper, Maldacena, Shenker and Stanford derived a temperature-dependent `bound to chaos' \cite{Maldacena2016bound}, formulated in terms of a Lyapunov exponent expressed as a function of out-of-order correlators containing correlations between {\it four}   matrix elements. 

Considering the matrix elements of an operator in the eigenbasis of a {\it chaotic} Hamiltonian as a random variable is the subject of the Eigenstate-Thermalization Hypothesis, developed by Deutsch and Srednicki \cite{deutsch1991quantum,srednicki1994chaos} in the '90s.   
When these ideas were put to test in the context of the OTOC, it became clear that ETH as was known was incomplete, because it only dealt with two-element correlations, and neglected all those correlations that were essential for the computation of four-point functions.

The full form of ETH, afterwards developed in Ref.~\onlinecite{foini2019eigenstate}, is an ansatz for the $k$-point correlations:
 \begin{align}
     \label{ETH_full}
     \overline{A_{i_1i_2}A_{i_2i_3}...A_{i_ki_1} } &= e^{(1-k)S(e_+)} F_{e^+}^{(k)}(\omega_{i_1i_2},...,\omega_{i_{k-1}i_k}) \\
    & \text{for different indices } i_1, i_2, \dots i_k \nonumber
 \end{align}
  where $e_+ =  (E_{i_1}+...+E_{i_k})/k$, $\omega_{i_k i_{k+1}}= E_k-E_{k+1}$ are the energy differences and $S$ is the thermodynamic entropy. The smooth functions $F^{(k)}$ define the operator.
  This expression is for different indices, but what about the host of possibilities with
  different index repetitions that we are now forced to take into account?

\subsection*{Free Probability: the framework}
  It turns out that the treatment of all this proliferation of possibilities, plus the development of the combinatorics that ensues, had already been developed in the field of {\it Free Probability} introduced by Voiculescu \cite{voiculescu1992free} and greatly developed by Speicher \cite{speicher1997free} and many others.
  Free probability may be seen as generalizing the notion of independence for noncommutative variables.
  For our purposes, the main ingredient is the definition of `free cumulants' of operators, for example:
$\kappa_{2k}(A_1, A_2 ,A_3,  \dots, A_{2k} )$, a set of a fixed combination of traces of products, in the spirit of (and taking the place of) ordinary 
cumulants.   {\it The full ETH is expressed by saying that expectations of products with different indices are directly connected with the free cumulants} \cite{pappalardi2022eigenstate}.


 An important notion we will use here is:
 two operators $B$ and $A$ are said to be {\it free} (asymptotically, for large size) if their free cumulants vanish as follows:
 \begin{equation}
 \kappa_{2l}(A, B, A, B, \dots A, B) = 0 \quad \forall l\leq k\ . \label{chose}
 \end{equation}
Actually, the definition of freeness involves arbitrary mixed sequences of $A$ and $B$, for simplicity, we illustrate our arguments using alternating sequences.
Importantly, $B$ and a randomly rotated matrix $A=UA'U^\dag$ are asymptotically free when $U$ is taken to be uniformly distributed over the full unitary group. It is common knowledge that for large matrices, certain physical transformations bring the eigenvectors of $A'$ and $B$ to become ``generic'' or ``typical'' with respect to each other. The definition of freeness replaces this vague notion and provides the tools for computing mixed correlations from the individual asymptotic properties of $A$ and $B$.
In this work, we shall relax this condition by defining {\bf $k$-freeness} when (\ref{chose}) is true only up to some $k$.
A comment is in order. In most practical settings, rather than requiring the free cumulant Eq.~\eqref{chose} to vanish exactly, it is more convenient to require that its absolute value is smaller than some arbitrary precision $\epsilon$. For simplicity, in what follows we will always write ``$=0$'' as in Eq.~\eqref{chose}, although it is intended that there are small corrections, which vanish only asymptotically in the appropriate limit.

 \subsection*{$k$-designs}
A third development coming from quantum information is the problem of computing flat, structureless averages over the set of unitary matrices, which formally corresponds to the integration over the unitary group with the Haar measure. As already mentioned in the introduction, this notion has many practical applications in quantum information but, sampling unitaries directly from the Haar measure is unpractical, and it is convenient to find simpler ensembles that can approximate the uniform measure to some degree.\\
The basic idea had been introduced
in the context of integrals over the $d$-dimensional sphere $S^d$ \cite{delsarte1991spherical}: it is to find a set of points ($k$-design) such that for all polynomials of degree $k$ the averages
over the set coincide with the full flat average. 
Ref.~\cite{dankert2009exact} extended the notion of $k$-design to the unitary group $\mathrm{U}(D)$ ($D$ being the dimension) introducing \emph{unitary designs}. A set $\mathcal E$ of unitaries $U$ is a unitary $k$-design if the averages of all polynomials of degree $k$ of the matrix elements of $U$ taken over the set $\mathcal E$ and over all the Haar ensemble coincide.
This is a measure of how good our set $\mathcal E$ of unitaries is as a proxy for the whole group. 
The same definition can be given in terms of a different object:
the superoperator known as $k$-fold \emph{quantum channel}
$$ \Phi_\mathcal E^{(k)}(A_1
\otimes\dots \otimes  A_k) = \mathbb E_{U\sim \mathcal E}[U^{\dagger \, \otimes k} A_1\otimes \dots
\otimes A_k U^{\otimes k}] \ ,$$
which quantifies how the combined action of $U$ and $U^\dagger$ act on observables defined on $k$ copies of the Hilbert space, i.e. the $k$ replicas in the left side of Fig.\ref{fig_summary}.
Or equivalently, with summation convention:
\begin{eqnarray} & & \Braket{ a_1, \dots, a_k| \Phi_\mathcal E^{(k)}(A^1\otimes 
\dots \otimes A^k) | b_1,\dots,b_k} \nonumber \\
&=& \mathbb E [U^{\dagger}_{a_1 \bar a_1} 
\dots U^{\dagger}_{ a_k \bar a_k}   U_{ \bar b_1 b_1} \dots
 U_{ \bar b_k b_k} ] A^1_{ \bar a_1 \bar b_1} 
\dots  A^k_{\bar a_k \bar b_k}  \ .\end{eqnarray}
 In this language $\mathcal{E}$ is a $k-$design if $\Phi_\mathcal E^{(k)}$ coincides with the Haar $k$-fold channel $\Phi_{\rm Haar}^{(k)}$.

In this work, we will look at the properties of $k$-fold quantum channels {\bf asymptotically}, in the limit of large dimension $D$.

\subsection*{Main results}

In this work, we show how Free Probability provides new insights into the study of unitary designs and their connection to the eigenstate-thermalization hypothesis. This is based on the definition of { $k$-freeness} which is obtained by asking that the condition (\ref{chose}) is valid only up to some finite $k$.
The main results are pictorially illustrated in Fig.\ref{fig_summary} and can be summarized as follows:
\begin{enumerate}
    \item Unitary $k$-designs can be rephrased as asymptotic $k$-freeness for any operator. The relation between freeness and designs is the following: two operators transformed with respect to one another with a randomly chosen element of a $k$-design are asymptotically $k$-free.  Specifically, in the limit of large $D$
    \begin{align}
        \begin{split}
                    k-\text{design} \Longrightarrow \kappa_{2k}(A_1^U,B_1, \dots A_k^U,B_k) =0 \\
        \forall A_i, B_i\ , \text { up to } k \ ,
        \end{split}
    \end{align}
   where $A_i^U=U A_i U^{\dagger}$ and $U$ is sampled from a $k$-design. 
   As a consequence, we may use all the tools and combinatorics developed in the free probability literature to study designs. Particularly useful for quantum information are:
    \begin{itemize}
        \item[1.1] Formulas for computing $2k$-OTOC in terms of individual moments of $A_1,\dots, A_k$ and $B_1,\dots B_k$ separately;
        \item[1.2] Expressions for the $k$-fold quantum channel as an expansion for large $D$ in terms of free cumulants. 
    \end{itemize}
    \item ETH implies that unitary time evolution ${U(t) = e^{-i H t}}$ leads at infinite times to freeness for all finite $k$, but only when restricted to ETH physical operators $\mathcal A$. 
     This means that in the limit of long times, physical observables $A(t) = U^\dagger(t) A U(t)$ and $B$ become $k$-free, for all finite $k$, if $A$ and $B$ obey ETH.    
    Specifically,
    \begin{align}
    \begin{split}
        \text{ETH}_{\mathcal A} \xRightarrow[\text{large times} ]{} \kappa^\beta_{2k}(A_1(t),B_1, \dots A_k(t),B_k) =0 \\
        \forall A_i, B_i \in \mathcal A \ , \forall k \ .
    \end{split}
    \end{align}
    This applies to the canonical version of free cumulants at any inverse temperature $\beta$.
    \begin{itemize}
        \item [2.1] This result is rigorous upon infinite time-average $\lim_{ t_{\rm max}\to \infty}\frac 1{t_{\rm max}} \int_0^{t_{\rm max}} f(t) dt$.
        \item [2.2] We define a physical operator-dependent time-scale $t_k$, named \emph{free-k time}, for which the {$2k$\nobreakdash-th} thermal-free cumulant vanishes. 
    \end{itemize}
    Analogously to the Haar case, more general statements with more times are straightforward. For instance, the limit of large time of the OTOC $\langle A_1(t)B_1 \dots A_k(t) B_k\rangle_\beta$ has a simple expression in terms of thermal moments of $A_1, \dots A_k$ and $B_1,\dots B_k$ separately.  
    
    The definition of ($k$-)freeness requires the vanishing of all mixed cumulants.
    However, in this context the vanishing of alternating free cumulants is enough.
    This is known for Haar rotated matrices \cite{nica2006lectures}, for ETH it follows from the fact that small powers of ETH observables obey ETH.
\end{enumerate}

\section{Notions and notations}
\label{sec_notion}
This section covers the notation and the known results that serve as the basis for this work. We start in Sec.~\ref{sec_kchan} by introducing $k$-channels and we discuss Haar averages and $k$-designs in Sec.\ref{sec_haar}.  In Sec.~\ref{sec_free} we provide a small introduction to Free Probability. Finally, in Sec.~\ref{sec_ETH}, we discuss the full ETH and its relation with Free Probability. 

\subsection{$k$-fold quantum channels}
\label{sec_kchan}
Let $\mathcal H$ be a $D$-dimensional Hilbert space and $\mathcal U(\mathcal H)$ the set of unitary operators in $\mathcal H$.
We consider an arbitrary \emph{ensemble of unitaries} $\mathcal E=\{p_j, U_j\}$, where $p_j$ are probability distributions ($\sum_ip_i=1$) and $U_j$ are unitary operators. The ensemble average of a function $f(U)$ over $\mathcal E$ is given by
\begin{equation}
	\mathbb E_{U\sim \mathcal E}[f(U)]:= \sum_i p_i f(U_i)\ . 
\end{equation}
One may work as well with continuous ensembles $\mathcal{E}$, specified by a probability distribution  $d\mu(U)$ over the set of unitaries. In this case, it is intended that the definition above should be replaced by
\begin{equation}
\label{eq:expectation-with-measure}
    \mathbb E_{U\sim \mathcal E}[f(U)]:= \int d\mu(U) f(U).
\end{equation}

A natural quantity to study is how a given ensemble acts on the density matrices of the system. An equivalent point of view we will adopt focuses instead on the observables, in which case we think of the unitaries as acting on the observables in the Heisenberg picture. Adopting this second point of view, we define the quantum channel $\Phi_{\mathcal{E}}$ associated with the ensemble $\mathcal{E}$ through its action on observables $O\in \mathcal H$
\begin{equation}
	\Phi_{\mathcal{E}}(O) = \mathbb E_{U\sim \mathcal E}[U^{\dagger} O U].
\end{equation}

To further investigate the correlation property of the ensemble $\mathcal{E}$ we introduce $k$ copies of $\mathcal H$ -- often referred to as replicas.
In the replicated space we can think of a more general class of observables $O^{}$, defined as generic operators acting on the replicated space $\mathcal H^{\otimes k}$.
Given an ensemble $\mathcal{E}$, we can then introduce the \emph{$k$-fold quantum channel} $\Phi_\mathcal E^{(k)}(O)$ defined as
\begin{equation}
\label{eq:k-fold-channel}
	\Phi_\mathcal E^{(k)}(O) = \mathbb E_{U\sim \mathcal E}[U^{\dagger \, \otimes k} O U^{\otimes k}],
\end{equation}
i.e., first, each replica is evolved with the same unitary $U$ drawn from the ensemble; afterwards, the average over $U$ is performed.

The notion of $k$-fold quantum channel is extremely useful from a physical perspective in its own right. An example of a setting where the $k$-fold channel plays an important role is when $U$ describes the time-evolution of a noisy system for a given disorder realization. In this case, we might be interested in computing non-linear functions of the evolved system's density matrix ---e.g. the exponential of its R\'enyi entropies--- for a given disorder realization, and average over disorder only afterwards, as commonly done in the context of random unitary circuits~\cite{PhysRevX.7.031016, PhysRevX.8.021014, PhysRevX.8.021013, PhysRevX.10.031066} (see also e.g. Ref.~\cite{Fisher_2023} for a review), noisy fermion transport~\cite{Bauer_2017, PhysRevLett.123.110601, PhysRevX.9.021007,Bernard_2019, Bernard_2021a, Bernard_2021b, Bernard_2022, Hruza_2023, swann2023spacetime, bernard2023exact}, or the SYK model~\cite{Jian_2021, S_nderhauf_2019, agarwal2022emergent, stanford2022subleading}.
Another application closer to the following discussion is the computation of the average $2k$-OTOC (later defined in Eq.~\eqref{eq_kOTOC}), which can be simply computed in terms of the $k$-fold channel~\cite{roberts2017chaos, 10.21468/SciPostPhys.12.4.130, Pappalardi2023quantum}.

\subsection{The Haar channel and $k$-designs}
\label{sec_haar}
Among all choices of ensembles, one of the simplest ones, particularly relevant to quantum information, is the uniform ensemble. This is the ensemble induced by the Haar measure $d\mu_{\rm Haar}$ over the unitary group.
The Haar measure is uniquely characterized by the fact that it is left- and right-invariant under multiplication by a unitary and that it is normalized.
In other words, for any fixed $V\in\mathcal U(\mathcal H)$, the measure over $U$ satisfies
\begin{align}
    d\mu_{\rm Haar}(VU)= d\mu_{\rm Haar}(UV) &=  d\mu_{\rm Haar}(U)\\
	\int d\mu_{\rm Haar}(U) &=1.
\end{align}
respectively due to left-and-right invariance and normalization.
The Haar measure can then be used to define the Haar $k$-fold channel combining Eqs.~\eqref{eq:expectation-with-measure} and~\eqref{eq:k-fold-channel}:
\begin{equation}
	\Phi_{\rm Haar}^{(k)}(O) = \int d\mu_{\rm Haar}(U) \; U^{\dag \, \otimes k} O U^{\otimes k} \ .
\end{equation}

Due to its properties, the Haar channel is useful in many applications in quantum information \cite{chau2005unconditionally, renes2004symmetric,hayashi2005reexamination, iblisdir2006optimal, Emerson_2005, PhysRevLett.108.110503, PhysRevLett.120.050406, PhysRevA.99.052323, huang2020predicting, paini2019approximate, morris2020selective, knips2020multipartite, PhysRevLett.122.120505, Elben_2022, low2010pseudorandomness}.
In practice, however, many applications only utilize the Haar channels up to a finite value of $k$. Furthermore, in many practical setups, it is difficult to directly sample unitary gates from the Haar ensemble at large $D$. For these reasons, it is convenient to find different ensembles that can reproduce the properties of the Haar $k$-fold channel at least up to some finite $k$.

This has inspired the notion of $k$-design. An ensemble $\mathcal{E}$ is said to be a $k$-design if 
\begin{equation}
	\label{eq:k-design}
	\Phi_{\mathcal E}^{(k)}(O)=\Phi_{\rm Haar}^{(k)}(O).
\end{equation}
Intuitively, a unitary $k$-design is indistinguishable from the Haar ensemble by studying up to the $k$th moment of $\mathcal{E}$. That is, for any polynomial $p_k$ containing up to $k$-th powers of $U$ and $U^\dagger$, then
\begin{equation}
    \mathbb{E}_{U\sim \rm Haar} \left[ p_k(U) \right] = \mathbb{E}_{U\sim \mathcal{E}} \left[ p_k(U) \right].
\end{equation}
By this definition, if an ensemble is a $k$-design, then it is also a $(k-1)$-design and, in general, a $k'$-design for all $k'<k$.

Furthermore, the action of the Haar $k$-channel can be computed explicitly using the Schur-Weyl duality (see e.g. Sec.~2 of Ref.~\onlinecite{roberts2017chaos} for an elementary introduction). The underlying idea is that for any $O$, $\Phi_{\rm Haar}^{k}(O)$ commutes with any unitary operator of the form $V^{\otimes k}$ with $V\in \mathcal{U}(\mathcal{H})$. From this it can be shown that it needs to be a linear superposition of operators acting as permutations of the $k$ replicas in $\mathcal{H}^{\otimes k}$.

To be more explicit, we introduce a representation of the permutation group $S_k$ onto the $k$-replica Hilbert space $\mathcal{H}^{\otimes k}$. A permutation $\alpha\in S_k$, specified as $\alpha =(\alpha(1), \dots, \alpha(k))$, acts on $\mathcal{H}^{\otimes k}$ through the permutation operator $P_\alpha$ as follows
\begin{equation}
	P_\alpha\left(\ket{i_1}\otimes \ket{i_2}\otimes \cdots \otimes \ket{i_k}\right) =
    \ket{\alpha(i_1)}\otimes \ket{\alpha(i_2)}\otimes \cdots \otimes \ket{\alpha(i_k)}
\end{equation}
with $\ket{i_l}\in\mathcal{H}$ for $1\leq l\leq k$.
From Schur-Weyl duality, it can be shown that
\begin{equation}
    \label{eqWg}
	\Phi_{\rm Haar}^{(k)}(O) = \sum_{\alpha \beta \in S_k} \text{Wg}_{\alpha, \beta}(D)\,\text{Tr}(P_\beta O) \, P_{\alpha^{-1}}\ ,
\end{equation}
where $\text{Wg}_{\alpha, \beta}(D) = \text{Wg}_{\alpha^{-1} \beta}(D)$ is the Unitary Weingarten matrix, given by the inverse of the Gram matrix\footnote{{We assume the existence of the inverse $Q^{-1}$, valid for $k\leq N$, which is the case under consideration.}} \cite{collins2003moments, gu2013moments}
\begin{equation}
	Q_{\alpha, \beta} =\text{Tr}( P_\alpha^{-1} P_\beta) = D^{\#(\alpha^{-1}\beta)} \ ,
\end{equation}
and $\#\alpha$ counts the number of cycles in the permutation $\alpha$\footnote{
Note that this expression is the same as in Ref.\cite{roberts2017chaos} upon substituting $\alpha^{-1}$ with $\alpha$. Here we use $P_{\alpha^{-1}}$ for consistency with the standard notation for the Weingarten matrix \cite{collins2003moments}. }.
While the above result provides a direct way to compute $\Phi_{\rm Haar}^{(k)}(O)$ for finite $k$, the general expressions obtained in this way quickly become unwieldy as $k$ increases. One first simplification can be obtained by considering the large-$D$ limit.
For example, in this limit, the first three orders for $O=A^{\otimes k}$ read
\begin{subequations}
\label{eq_PhiHaar}
\begin{align}
\Phi_{\rm Haar}^{(1)}(A) & = \langle A \rangle\,  P_{(1)}
\\ 
\Phi_{\rm Haar}^{(2)}(A^{\otimes 2}) & = 
\left(\langle A\rangle ^2 + o(1) \right)  P_{(1,2)}\nonumber\\
&~~+
\left( \langle A^2\rangle -  \langle A\rangle^2 + o(1) \right) \, \frac{P_{(2,1)}}{D}
\end{align}    
\begin{align}
\label{Phi3_}
	&\Phi_{\rm Haar}^{(3)}(A^{\otimes 3}) = \left(\langle A\rangle^3 +o(1) \right) P_{(1,2,3)}\nonumber\\
    &~+ 
	\left[ \langle A \rangle ( \langle A^2\rangle -  \langle A\rangle^2) + o(1) \right]\frac{ P_{(2,1,3)}+ P_{(1,3,2)} + P_{(3,2,1)}}{D}\nonumber\\
    &~+\left[\langle A^3\rangle -3 \langle A\rangle \langle A^2\rangle + 2\langle A\rangle^3 +o(1) \right] \frac{ P_{(2,3,1)}+ P_{(3,1,2)}}{D^2} 
\end{align}
\end{subequations}
where $o(1)$ denotes a vanishing number for $D\to \infty$.

We have also defined the normalized expectation for $A\in \mathcal H$ %
\footnote{We are furthermore implicitly assuming that while we take the limit $D\to\infty$, $\langle A^n \rangle$ remains finite. This will indeed be the case for many choices of operators in a many-body physics setting, i.e. this is true if $\langle A\rangle$ is a few-body or a local operator.}
\begin{equation}
\langle A \rangle =  \frac {\text{Tr}(A)} D	 \ ,
\end{equation}
 i.e. the expectation value of $A$ in the fully-mixed state.


Before proceeding we remark that in the expansion in Eq.\eqref{eq_PhiHaar} above, we neglected subleading corrections to the coefficient of each permutation operator, but retained all permutation operators, independently of the power of $D$ suppressing their effect. As we shall see in the next section, this turns out to be the correct way of expanding the output of the channel for our purposes, as each permutation operator will end up contributing an $O(1)$ amount to, e.g., the $k$-OTOC.

From the above expression we also begin to see that each $k$-fold channel contains information about $A$ not present in the $(k-1)$-fold channel. This new information appears in the last term of each of the expressions above, i.e. as the coefficient of the non-trivial cyclic permutations. To better understand this structure for arbitrary $k$ we will turn to Free Probability.

\subsection{Free Probability}
\label{sec_free}

\begin{figure*}
    \centering
    \includegraphics[width=\linewidth]{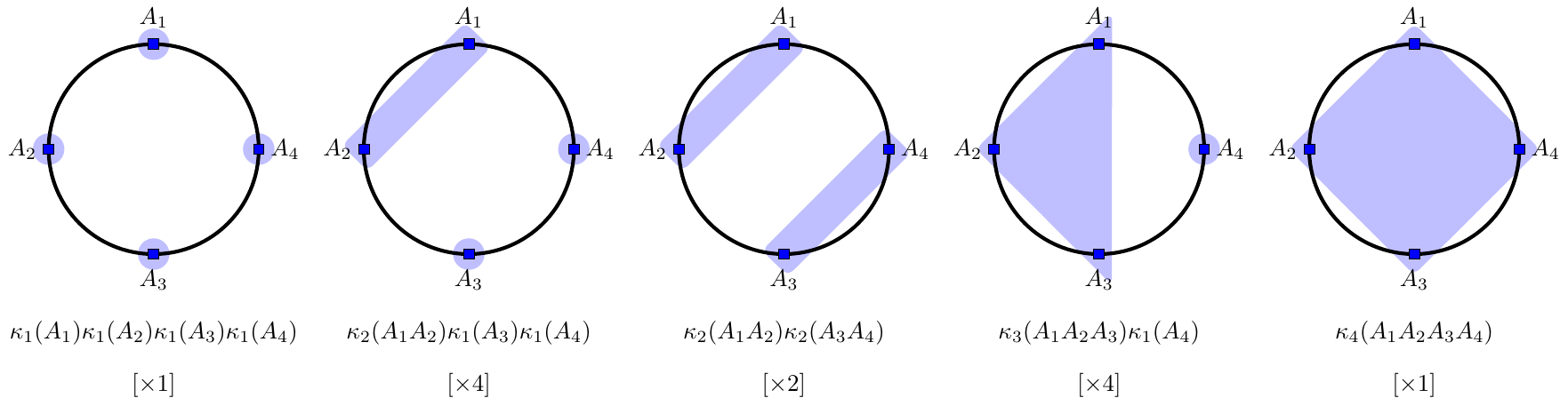}
    \caption{Graphical representation of five non-crossing partitions appearing in the expansion of $\langle A_1 A_2 A_3 A_4 \rangle$ and their relation to free cumulants through Eqs.~\eqref{eq_free_cumu_def} and ~\eqref{eq:free_cumulant_subdef}. In every picture, each shaded area correspond to a block of the partition encompassing a subset of $\{1,2,3,4\}$. In fact, every non-crossing partition of a set of four elements can be obtained by a rotation of one of the partition drawn above. The number $[\times n]$ written below the partition indicates how many distinct non-crossing partitions can be obtained in this way. Note that, since the partitions are non-crossing, terms like $\{\{1,3\},\{2,4\}\}$ are not included, as they would induce a crossing of the blocks in the plane. Through Eq.~\eqref{eq:free_cumulant_subdef} each non-crossing partition $\pi$ is associated with $\kappa_\pi$ a product of free cumulants $\kappa_{l\in\mathbb{Z}}$ over the blocks. This is explicitly reported below the picture of every partition.}
    \label{fig:free-cumulants}
\end{figure*}

Free probability -- or ``non-commutative probability'' -- can be thought of as the generalization of classical probability to non-commutative random variables, where the concept of ``freeness'' extends the one of ``independence''.  Let us discuss some basic properties, for all the details we refer to Refs.\cite{ nica2006lectures, speicher2009free, mingo2017free}.\\

For this purpose, we consider two random matrices $A$ and $B$, which could be drawn from different ensembles.
We then introduce the expectation value 
\begin{equation}
    \label{eq_defineExp}
    \langle\bullet\rangle_{\mathcal E} = \lim_{D\to \infty} \frac 1D 
    \mathbb E_{\mathcal E} \left [ \text{Tr}(\bullet) \right ]\ ,
\end{equation}
where here ${ \mathcal E} $ represents a generic random matrix ensemble \footnote{ In the literature of Free Probability, the expectation value is usually noted by $\phi(\bullet)$ \cite{speicher2009free}}. Note the slight abuse of notation, since we also denote $\langle\bullet\rangle = \Tr(\bullet)/D$ when the argument does not involve any random matrix. 
Before proceeding we comment on a technical assumption: the ensembles of $A$ and $B$ must be defined as a function of $D$, this can be done naturally in the context of many-body physics, where local (or few-body) operators can be naturally extended to act on a larger Hilbert space.
In free probability, it is further required that the moments $\braket{A^k}_{\mathcal E}$ exist for all $k$ (and similarly for $B$). Again, this assumption is naturally satisfied for local (or few-body) operators embedded in a many-body system.
Finally, an operator with zero expectation value is said to be centered.

In terms of the expectation value above, we can define the notion of freeness for $A$ and $B$. These are said to be free if~\cite{voiculescu1991limit, voiculescu1992free}
\begin{multline}
    \label{eq_free_AB}
	\bigg\langle \big(A^{n_1}-\langle A^{n_1} \rangle _{\mathcal E}\big)\, \big(B^{m_1}-\langle B^{m_1} \rangle _{\mathcal E} \big)  \cdots\\
    \times \big(A^{n_k}-\langle A^{n_k}\rangle _{\mathcal E} \big) \big(B^{m_k}-\langle B^{m_k}\rangle _{\mathcal E}\big)
	\bigg\rangle_{\mathcal E}  = 0
\end{multline}
for all $n_1,\dots  m_k \geq 1$.

The product of the terms above is said to be \emph{alternating}, since consecutive powers of neither $A$ nor $B$ appear.
The definition of freeness can then be enunciated as: ``$A$ and $B$ are free \emph{if the alternating product of centered elements is centered}''. The definition can be extended to different sequences of matrices $\{A_1, A_2, \dots\}$ and $\{B_1, B_2, \dots \}$. Let us remark that the definition of freeness always depends on the expectation value $\langle \bullet 
\rangle_{\mathcal E}$, cf. Eq.\eqref{eq_defineExp}. 

This definition is not particularly transparent, but it has to be understood as a rule for computing mixed moments of non-commuting variables from knowledge of the moments of individual variables. For instance, if $A$ and $B$ are two free random matrices, we can simplify the expectation value of the product
\begin{equation}
\label{eq_ABAB}
\begin{split}
	 \langle ABAB \rangle_{\mathcal E}  = \langle A^2 \rangle_{\mathcal E}  \langle B\rangle^2_{\mathcal E} +  \langle B^2\rangle _{\mathcal E} \langle A\rangle_{\mathcal E} ^2 - \langle B\rangle^2 \langle A\rangle_{\mathcal E} ^2  \ ,
\end{split}
\end{equation}
as can be seen by expanding the products and setting to zero all alternating products, e.g. using that
\be
    \bigg\langle (A - \langle A\rangle_{ \mathcal E}) (B - \langle B\rangle_{  \mathcal E}) (A - \langle A\rangle_{  \mathcal E}) (B - \langle B \rangle_{  \mathcal E})\bigg \rangle_{\mathcal E}  = 0.
\ee
The definition of freeness can be rephrased in a compact way using the concept of free cumulants and the combinatorial theory of freeness, based on non-crossing partitions, as developed by Speicher \cite{speicher1997free}.
In the following we will first define non-crossing partitions. In terms of these, we will introduce the notion of free cumulants, and finally we will show how expressing expectation values in terms of free cumulants simplifies the definition of freeness~\eqref{eq_free_AB}.

A partition of a set $\{1, \dots n\}$ is a decomposition in blocks that do not overlap and whose union is the whole set. To define non-crossing partitions we can proceed as described in Fig.~\ref{fig:free-cumulants} for $n=4$. We fix an order, by arranging $n$ points on a circle and depicting each block as a polygon (including the degenerate cases of one-vertex and two-vertices polygons) whose vertices are a subset of the $n$ points. Partitions in which ``blocks do not cross'', i.e., polygons do not overlap, are called \emph{non-crossing partitions}. The set of all non-crossing partitions of $\{1, \dots n\}$ is denoted $NC(n)$.  The \emph{free cumulants} $\kappa_n$ are defined implicitly from the moment-cumulant formula, which states that the moments of variables $A_i$ read
\begin{align}
    \label{eq_free_cumu_def}
    \langle A_{1} \dots A_{n} \rangle_{\mathcal E}  &:= \sum_{\pi \in NC(n)} \kappa_{\pi} (A_{1}, \dots, A_{n} ) 
    \qquad
    \\
    \label{eq:free_cumulant_subdef}
    \kappa_\pi (A_{1}, \dots, A_{n} ) &:= 
 \prod_{\beta\in\pi} \kappa_{|\beta|} ( A_{j_1},\dots ,A_{j_{|\beta|}} )
\end{align} 
where $\beta = (j_1, \dots, j_\beta)$ denotes the element of a block of the non-crossing partition $\pi$, with $|\beta|$ numbers of elements. 
See Fig.\ref{fig:free-cumulants} for a graphical illustration. 

Note that the formula above shall indeed be used as a recursive definition of $\kappa_n$ since all the other terms in Eq.~\eqref{eq_free_cumu_def} are defined by imposing that the same definition holds for $n-1$.
For example, in the simple case where $A_{i}=A$, the $\kappa_n$ for $n\leq4$ is given by
\begin{subequations}
\label{eq_k4}
	\begin{align}
		\kappa_1(A) & = \langle A \rangle_{\mathcal E} \ ,
		\\
		\label{k2}
		\kappa_2(A, A) & = \langle  A^2 \rangle_{\mathcal E}  -  \langle  A \rangle_{\mathcal E}   \langle  A \rangle_{\mathcal E} \ ,
		\\
		\kappa_3(A,A,A) & =  \langle A^3\rangle_{\mathcal E}  +2   \langle   A \rangle^3_{  \mathcal E}-  3 \langle   A\rangle \langle  A^2\rangle_{\mathcal E} 
		\ ,
		\\
		\kappa_4(A,A,A,A) & = \langle A^4 \rangle_{\mathcal E}  - 2\langle A^2 \rangle^2_{\mathcal E}  - 4 \langle A \rangle_{\mathcal E} \langle A^3\rangle_{\mathcal E}  \nonumber\\
        &~~ + 10\langle A \rangle^2_{\mathcal E} \langle A^2 \rangle_{\mathcal E}  - 5 \langle A \rangle^4_{\mathcal E} \ .
	\end{align}
\end{subequations}
Let us stress that the definition of free cumulants depends on the expectation value $\braket{\bullet}_{\mathcal E}$, which in the case of random matrices involves an average over the ensemble.
Before proceeding, we point out that, if the sum in Eq.~\eqref{eq_free_cumu_def} were to run over all partitions, both crossing and non-crossing, then we would recover the standard definition of cumulants. As a consequence of this, standard cumulants and free cumulants coincide for $n<3$, as all partitions are non-crossing in this case. 

The inversion of Eq.~\eqref{eq_free_cumu_def} can be made systematic using combinatorial tools \cite{nica2006lectures}, leading to the so-called M\"obius inversion formula:
\begin{equation}
	\label{eq_moebius}
	 \kappa_{\pi} (A_{1}, \dots, A_{n} ) = \sum_{\sigma\in NC(n), \sigma \leq \pi} \,  \langle A_{1}, \dots, A_{n} \rangle_{\sigma} \, \mu(\sigma, \pi)\ ,
\end{equation}
where $\langle \bullet \rangle_{\sigma}$ is the product of moments, one for each term of the partition $\sigma$, viz.
\begin{equation}
    \label{eq_moment_bloch}
    \langle A_1, \dots, A_n \rangle_\sigma := \prod_{\beta\in\sigma} \big\langle  A_{j_1} \dots A_{j_\beta}\big\rangle_{\mathcal E}
\end{equation}
and $ \mu(\sigma, \pi)$ is the so-called M\"obius function and ${\sigma\leq \pi}$ indicates that the sum is restricted to the partitions where each block of $\sigma$ is contained in one of the blocks of $\pi$.

\begin{figure*}
    \centering
    \includegraphics[width=\linewidth]{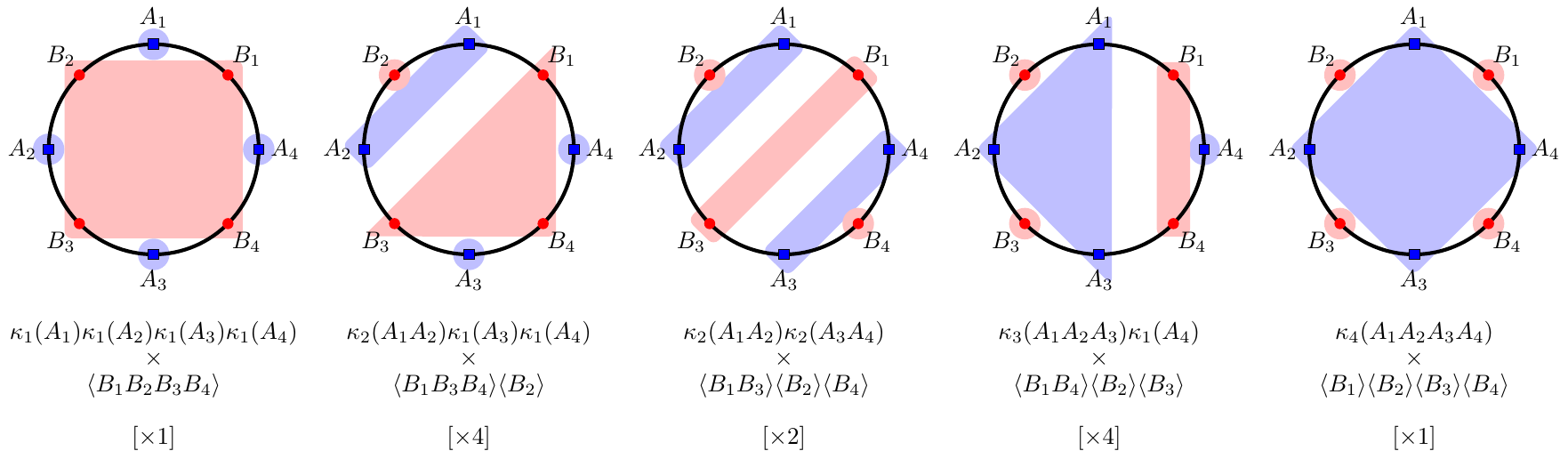}
    \caption{Graphical illustration of some of the terms appearing in Eq.~\eqref{eq_free_prod} for $n=4$. The shaded blue areas denote non-crossing partitions of the set $\{A_1,A_2,A_3,A_4\}$, as in Fig.~\ref{fig:free-cumulants}. To each non-crossing partition $\pi$, one can associate a dual non-crossing partition $\pi^*$ on the set $\{B_1,B_2,B_3,B_4\}$, denoted here as shaded red areas. The contributions of these terms to the sum in Eq.~\eqref{eq_free_prod} is the product of two terms reported below each partition. With $[\times n]$ we indicate that there are $n$ arrangements of that non-crossing diagram. }
    \label{fig:complement}
\end{figure*}

A salient property of free cumulants is that they characterize freeness by the vanishing of the \emph{mixed}
\footnote{Mixed means that the random variables appearing in the free-cumulant do not all come from the same sub-algebra.} free cumulants, as shown e.g. in Ref.~\cite{mingo2017free}.
{
 In the case of large random matrices, one has that two families $\{A_1, A_2, \dots  \}$ and $\{B_1, B_2,\dots\}$ are \textbf{asymptotically free} if all mixed cumulants are zero: 
{ \begin{align}
 \begin{split}
\kappa_{n}(a_1, a_2, \dots, a_{n}) = 0 \quad \forall n\ ,
\end{split}
\end{align}
 where $a_1, \dots, a_n$ are letters of $\{A_1,A_2 \dots\}$ and $ \{B_1, B_2,\dots\}$, containing at  least a pair $(A_i, B_j)$ for some $i, j$.}
 in particular for any $n$ and for any set of indices $i_1,\dots,i_n$, and $j_1,\dots,j_n$
 \begin{align}
\begin{split}
       \label{eq_free_k}
	{\text{if }}\{A_1, A_2, \dots  \}\text{ and } \{B_1, B_2,\dots\}  \text{ are asymptotically
    free,}
    \\
    \text{then }
\kappa_{2n}(A_{i_1}, B_{j_1}, A_{i_2}, B_{j_2}, \dots A_{i_n} ,B_{j_n}) = 0 \quad \forall n\ .
\end{split}
\end{align}
}
This works as a rule for the computation of mixed moments (as in Eq.\eqref{eq_ABAB}) in terms of moments in which $A_1, \dots, A_n$ and $B_1, \dots, B_n$ appear separately. Even more nicely, one can find an explicit expression that relates the two.  Concretely, if $A_1, \dots A_n$ and $B_1, \dots B_n$ are asymptotically free, then the mixed moments are given by \cite{mingo2017free}
\begin{multline}
\label{eq_free_prod}
\langle A_1 B_1 A_2 B_2 \dots A_n B_n\rangle_{\mathcal E} = \\
=\sum_{\pi \in NC(n)} \kappa_\pi(A_1, \dots, A_n) \, \langle B_1, \dots, B_n\rangle_{\pi^*} \ ,
\end{multline}
where $\pi^*$ is the dual of the partition $\pi$ (also known as Kraweras complement), which is defined below.
Graphically (see Fig.~\ref{fig:complement} for $n=4$), the blocks composing $\pi^*$ are the maximal blocks (polygons) with vertices on  ``$B$'' that do not cross the blocks of $\pi$.
As a simple check, one can verify that for the simple case of $\langle ABAB\rangle_{\mathcal E} $, the general result~\eqref{eq_free_prod} immediately yields Eq.\eqref{eq_ABAB}.

While the right-hand side of Eq.~\eqref{eq_free_prod} is not manifestly symmetric under $A\leftrightarrow B$, one can also exchange the role of $A$ and $B$. 
In this case one obtains 
\begin{multline}
    \langle A_1 B_1 A_2 B_2 \dots A_n B_n\rangle_{\mathcal E} \\
    = \sum_{\pi \in NC(n)} \langle A_1, \dots, A_n\rangle_{K^{-1}(\pi)} \, \kappa_{\pi} (B_1, \dots, B_n) \ ,
\end{multline}
where $K^{-1}$ is the inverse of the duality transformation $\bullet^*$, i.e. $K^{-1}(\pi^*)=\pi$\footnote{Formally $K^{-1}$ is distinct from the duality transformation $^*$. In fact, applying the duality twice, one would generate a cyclic permutation, corresponding to a clockwise rotation by two units in the notation of Fig.~\ref{fig:complement}.}.

Thus the definition of freeness in Eq.\eqref{eq_free_AB} is greatly simplified when expressed in terms of cumulants as in Eq.\eqref{eq_free_k}. First of all, one gets rid of the condition of ``centerness'' for the variables and relaxes the condition of ``alternating'' to ``mixed''. Secondly, there are several properties of free variables which become particularly easy. For instance, the free cumulants of the sum of two-free variables are give by the sum of the free cumulants, e.g.
\begin{equation}
	\label{eq_free_sum}
    \kappa_n\left(A+B\right) = \kappa_n\left(A\right) + \kappa_n\left(B \right)\ ,
\end{equation}
where, for ease of notation, we have denoted $\kappa_n(X, \dots, X)$ with $X$ appearing $n$ times as $\kappa_n(X)$.
These nice properties translate over the generating functions, such as the $R$-transform or the $S$-transform, see e.g. Ref.\cite{mingo2017free}.
We will, however, not deal with these topics in the present manuscript.

While the notion of freeness might seem exotic, many well-known distributions of random matrices produce free matrices asymptotically in the $D\to\infty$ limit.
For example,
independent Gaussian random matrices~\cite{voiculescu1991limit} are free w.r.t. each other, and Wigner matrices are free w.r.t. deterministic ones~\cite{mingo2017free} --- in both cases the statement holds asymptotically in the $D\to\infty$ limit.
Another instance of free random matrices that will be relevant for us is the following. Given two non-random (deterministic) matrices $A$ and $B$ (such that again $\braket{A^n}$ is finite for all $n$ and similarly for $B$) and given $U\sim\text{\rm Haar}$, then ~\cite{voiculescu1991limit}

\begin{equation}
    U^\dag A U \quad\text{and} \quad B  \quad \text {are asymptotically free as} \quad D\to\infty.
\end{equation}
 
This theorem says that unitarily invariant random matrix models are asymptotically free from deterministic matrices. As a consequence, all the relations of Eqs.\eqref{eq_free_k}-\eqref{eq_free_prod} hold. The eigenvalue distribution is not changed by the random rotation that, however, changes the relationship between the eigenvectors of $A$ and $B$ which we would have called ``generic'' or ``typical'' and we can now refer to as ``free'.

\subsection{The full Eigenstate Thermalization Hypothesis}
\label{sec_ETH}
In a many-body system of $N$ particles described by a Hamiltonian $H$, the information about higher-order correlations is encoded in   
equilibrium correlators depending on $n$ times, i.e.
\begin{equation}
\label{eq_Aq}
    \langle A(t_1) A(t_2) \dots A(t_n) \rangle^\beta \quad \quad \text{with}\quad \langle \bullet \rangle^\beta = \frac 1Z \Tr \left ( e^{-\beta H} \bullet \right ) 
\end{equation}
and $ Z=\Tr \left ( e^{-\beta H} \right )$.
In principle, correlation functions at order $n$ contain new information that is not encoded in lower moments, which implies that matrix elements $A_{ij}=\bra{E_i} A|E_j\rangle$ of physical observables in the energy eigenbasis need to be correlated.
This lead Ref.~\cite{foini2019eigenstate} to introduce a full version of the Eigenstate-Thermalization-Hypothesis, as an ansatz on the statistical properties of the product of $n$ matrix elements $A_{ij}$. Specifically, averages of products with distinct indices $i_1\neq i_2 \dots \neq i_n$ read
\begin{equation}
    \label{ETHq}
    \overline{A_{i_1i_2}A_{i_2i_3}\dots A_{i_{n}i_1}} = e^{-(n-1)S(E^+)} F_{e^+}^{(n)}(\omega_{i_1i_2}, \dots, \omega_{i_{n-1}i_n}) 
\end{equation}
while, with repeated indices, it factorizes in the large $D$ limit as
\begin{multline}
\label{ETH_conta}
 \overline{A_{i_1i_2}\dots A_{i_{k-1}i_1}A_{i_1i_{k+1}}\dots A_{i_{n}i_1}} \\
 =  \overline{A_{i_1i_2}\dots A_{i_{k-1}i_1}} \; 
\overline{A_{i_1i_{k+1}}\dots A_{i_{n}i_1}}    \ .    
\end{multline}
In Eq.\eqref{ETHq}, $Ne^+ = E^+=(E_{i_1}+\dots +E_{i_n})/q$ is the average energy, $\vec 
\omega = (\omega_{i_1i_2}, \dots, \omega_{i_{n-1}i_n})$ with $\omega_{ij}=E_i-E_j$ are $n-1$ energy differences and $F_{e^+}^{(n)}(\vec \omega)$ is a smooth function of the energy density $e^+=E^+/N$ and $\vec \omega$. Thanks to the explicit entropic factor, the functions $F^{(n)}_e(\vec \omega)$ are of order one and they contain all the physical information. 
For $n=1, 2$ one retrieves the standard ETH ansatz \cite{srednicki1999approach}, i.e. $F^{(1)}_e=\mathcal A(e)$ is the microcanonical diagonal value, while $F^{(2)}_e=|f(e, \omega)|^2$ encodes dynamical correlation functions. 

The relation between the full ETH ansatz \eqref{ETHq}-\eqref{ETH_conta} and the equilibrium correlators Eq.\eqref{eq_Aq} becomes more apparent by the use of Free Probability \cite{pappalardi2022eigenstate}. Revisiting the definition \eqref{eq_free_cumu_def}, thermal free cumulants $\kappa^\beta_n$ are defined implicitly by
\begin{multline}    
    \langle A(t_1) A(t_2) \dots A(t_n) \rangle^\beta \\
    = \sum_{\pi\in NC(n)} \kappa^\beta_\pi \left ( A(t_1), A(t_2), \dots,  A(t_n) \right )  \ ,
\end{multline}
where $\kappa^\beta_\pi$ are products of thermal free cumulants one for each block of $\pi$. Exactly as in Eq.\eqref{eq_free_cumu_def}, this is just an implicit definition of cumulants in terms of moments, but the ETH ansatz \eqref{ETHq}-\eqref{ETH_conta} implies a particularly simple form for them at the leading order in $N$. Namely, the thermal free cumulants of ETH-obeying systems are given only by summations with distinct indices \cite{pappalardi2022eigenstate}
\begin{widetext}
\begin{align}
    \begin{split}
    \label{freekETH}
    \kappa^{\beta}_{n} \left (  A(t_1), A(t_2), \dots A(t_n) \right )  & = 
    \frac 1Z \sum_{i_1\neq i_2 \neq \dots \neq i_n} e^{-\beta E_i} A_{i_1i_2}A_{i_2i_3}\dots A_{i_{n}i_1}e^{i t_1 \omega_{i_1i_2} +\dots i t_n \omega_{i_{n}i_1}}\ .
    \end{split}
\end{align}
\end{widetext}
This result shows that all the non-gaussianities of the full ETH \eqref{ETHq} are encoded precisely in the thermal free cumulants. Explicitly, their Fourier transform $\text {FT} [\bullet]= \int d\vec \omega e^{i \vec \omega \cdot \vec t} \bullet$ reads
\begin{equation}
\label{eq:free-cumulant-ETH-intro}
      \kappa^{\beta}_{n} \left (  A(t_1), A(t_2), \dots ,A(t_n) \right ) = \text {FT} \left [ F^{(n)}_{e_\beta}(\vec \omega) e^{-\beta \vec \omega \cdot \vec \ell_n} \right ] 
\end{equation}
where the thermal weight with $\vec \ell_n = \left (\frac{n-1}{n}, \dots, \frac 1n , 0\right )$ corresponds to a generalization of the fluctuation-dissipation theorem \cite{haehl2017thermal}. The validity of the full ETH ansatz and its free cumulants decomposition~\eqref{freekETH} has been recently tested numerically in chaotic local many-body systems in Ref.~\cite{pappalardi2023general}. Conversely, in Bethe-ansatz-integrable models, the structure of the statistics of the matrix elements of operators is more involved, as recently explored in Ref.~\cite{essler2023statistics}.

\section{From $k$-design to $k$-freeness}
\label{sec_result_des}

 We now present the results in the context of sets of $D \times D$ matrices and the full unitary group, where there is a solid mathematical basis.  In the section below, we will extrapolate them to the quantum context, for which the rotational invariance is only local.

Let us rephrase the properties of $k$-designs in terms of Free Probability. We start by considering the case of only two operators $A$ and $B$ on $\mathcal H$.
As discussed in Sec.\ref{sec_free}, if $U$ is sampled according to the Haar distribution, then $A^U=U^\dagger A U$ and $B$ are asymptotically free \cite{voiculescu1991limit}. This means that $\forall k$ mixed cumulants vanish \cite{speicher2009free}
\begin{equation}
     \kappa_{2k}(A^U,B, \dots A^U, B ) = 0\ ,
\end{equation}
where we recall the ensemble average over $U$ is already included in the definition of the free-cumulants \eqref{eq_free_cumu_def} via Eq.\eqref{eq_defineExp}. 
It now follows that \emph{if $U$ is sampled from a $k$-design, then the mixed free-cumulants vanish up to order $k$}, in other words, $A^U$ and $B$ are {\bf asymptotically $k$-free}.

As such, they are characterized by the properties that characterize free variables up to order $k$, but not above. In particular, if $\Phi_{\mathcal E}(A^{\otimes k}) = \Phi_{\rm Haar}(A^{\otimes k}) $, one has:
\begin{enumerate}
    \item The vanishing of the $2k$-th free cumulant 
    \begin{equation}
    \label{mainB}
     \kappa_{2k}(A^U, B, \dots, A^U, B ) = 0\ .
\end{equation}
This should be seen as an alternative and straightforward way to check if an ensemble is a $k$-design directly from a connected correlation of some observables.
    \item Mixed $2k$-OTOC can be directly computed from the individual moments of $A$ and $B$ from
    \begin{equation}
        \label{mainC}
      \left\langle A^U B A^U B \dots A^U B \right\rangle_{\rm Haar}
      = \sum_{\pi \in NC(k)} \kappa_\pi(A^k) \, \langle B^k \rangle_{\pi^*}  \ ,
    \end{equation}
    hence providing a direct generalization to arbitrary $k$ of the results of \cite{roberts2017chaos}.
    \item In the large $D$ limit, the $k$-fold Haar channel can be decomposed in terms of free cumulants:
\begin{equation}
\label{eq:mainA_a}
	\Phi_{\rm Haar}^{(k)}(A^{\otimes k}) = \sum_{\alpha \in S_k} \frac{P_{\alpha^{-1}}}{D^{k-\#\alpha}} \left( \kappa_\alpha(A, \dots, A) + o(1)\right)\ ,
\end{equation}
where $\kappa_\alpha(A, \dots, A)$ are related to the free cumulants $\kappa_\pi$.  This expression extends Eqs.\eqref{eq_PhiHaar}  to general $k$ and it should be understood as an expansion in large $D$ for the coefficients of each permutation operator. More precisely, the subleading corrections to the coefficients, denoted as $o(1)$, scale like $O(D^{-2})$ for large $D$.
\end{enumerate}

\begin{figure*}
    \centering
    \includegraphics[width=\linewidth]{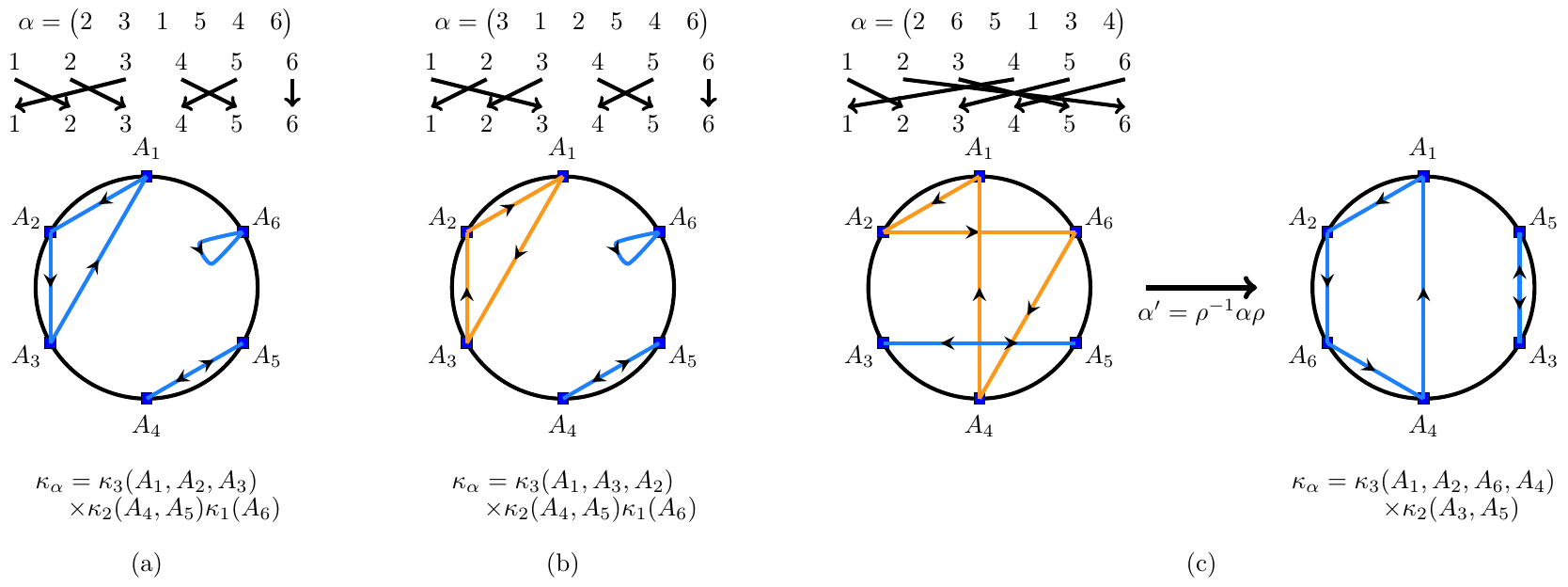}
    \caption{Graphical illustration of some of the coefficients $\kappa_\alpha$. Each panel reports a permutation of $6$ elements. The top row reports $\alpha$ both in the form $\alpha=\begin{pmatrix}
        \alpha(1)&\alpha(2)&\dots
    \end{pmatrix}$ and in a more graphical form as well. Below this, the permutation is drawn as a set of arrows on the circle (c.f. Fig.~\ref{fig:free-cumulants}). Panel (a) reports a permutation that satisfies both conditions (i) and (ii) ---see main text--- therefore $\kappa_\alpha$ can be directly computed as the product of free cumulants over the orbits. Panel (b) reports a permutation that satisfies condition (i), but not (ii), since the cycle in green is clockwise. The coefficient $\kappa_\alpha$ can be obtained as the product of free cumulants, but the ordering of matrices must be consistent with the order of the cycle. Panel (c) reports a more general permutation. The coefficient $\kappa_\alpha$ in this case can be computed by reordering the elements in such a way that after reordering, conditions (i) and (ii) are satisfied.}
    \label{fig:permutations}
\end{figure*}

In what follows, we build some intuition about these results by discussing them 
in the general case of multiple operators $A_1, A_2, \dots A_k$ and $B_1, B_2,\dots, B_k$. 
From the point of view of channels, this implies the results apply to arbitrary $A_1\otimes A_2\dots \otimes A_k$ and $B_1\otimes B_2\dots \otimes B_k$.\\

\subsection{Haar $k$-fold channel and free cumulants}
\label{sec_haarkfold}

The structure of $\Phi^{(k)}_{\rm Haar}$ can be highly simplified using free-cumulants. At the leading order in $D$ we find 
\begin{equation}
\label{mainA}
	\Phi_{\rm Haar}^{(k)}(A_1\otimes \dots A_k) 
= \sum_{\alpha \in S_k} \frac{P_{\alpha^{-1}}}{{D^{k-\#\alpha}}} \left ( {\kappa_\alpha(A_1,\dots ,A_k)} + o(1)\right )
\end{equation}
where we recall that $\#\alpha$ is the number of cycles of $\alpha$.

While the derivation of this expression can be found in App.\ref{app_deri}, let us discuss here that the coefficients
$\kappa_\alpha$  can be most easily expressed in terms of free cumulants as follows. \\
A permutation $\alpha\in S_n$ naturally induces a partition $\pi$ of $n$ elements through the set of its orbits $\pi=\text{Orb}(\alpha)$. An orbit ---also referred to as cycle in the context of permutations--- is a subset of the $n$ elements that can be obtained by repeatedly acting with $\alpha$ on a certain element, e.g. $\{ 1, \alpha(1), \alpha(\alpha(1)), \dots \}$ is an orbit. Varying the initial element, different subsets can be obtained, inducing a partition over $\{1,\dots,n\}$. We begin by considering a permutation that satisfies the following two conditions: 
(i) Within each of its orbits, the elements are permuted along the direction specified by the cyclic permutation $\gamma = (2, 3, \dots, n, 1)$; in practice, w.r.t. our graphical notation where the elements are ordered in the counterclockwise direction, this means that, within each cycle, the permutation acts in the counterclockwise direction (ii)
the associated partition $\pi$ formed by its orbits is non-crossing. 
In Fig.~\ref{fig:permutations}(a) and (b), we respectively provide an example of a permutation satisfying and violating condition (i). If $\alpha$ satisfies both conditions, then
\ba
\label{eq:kappa-alpha-kappa-pi}
    \kappa_\alpha(A_1,\dots ,A_k) &= \kappa_\pi(A_1,\dots ,A_k).
\end{align}

On the other hand, if $\alpha$ does not satisfy either of the conditions, this is simply due to the arbitrary ordering given to the operators $A_1,\dots,A_k$ in the tensor product. Therefore, one can simply re-order the operators $A_1,\dots,A_k$ in such a way that the partition $\alpha$ satisfies conditions (i) and (ii) after reordering. Formally, this is due to the fact that for any permutation $\rho$
\be
\label{eq:permutation-conjugation}
    \kappa_{\alpha}(A_1,\cdots, A_k) = \kappa_{\rho^{-1} \alpha \rho} (A_{\rho(1)},\cdots, A_{\rho(k)}),\, \forall \rho \in S_k
\ee
and, for a given $\alpha$, $\rho$ can be chosen in such a way that, on the right-hand side, $\rho^{-1} \alpha \rho$ satisfies (i) and (ii). The computation of $\kappa_\alpha$ can therefore always be reduced to the evaluation of free cumulants through Eqs.~\eqref{eq:kappa-alpha-kappa-pi}~\eqref{eq:permutation-conjugation}. See Fig.~\ref{fig:permutations}(c) for an example of such reordering.\\

We have thus expressed Haar channels as a sum of free cumulants as defined in Eq.\eqref{eq_free_cumu_def}.
In the first few orders, this reads

 \begin{widetext}
\begin{figure}[H]
    \includegraphics[width=1\textwidth]{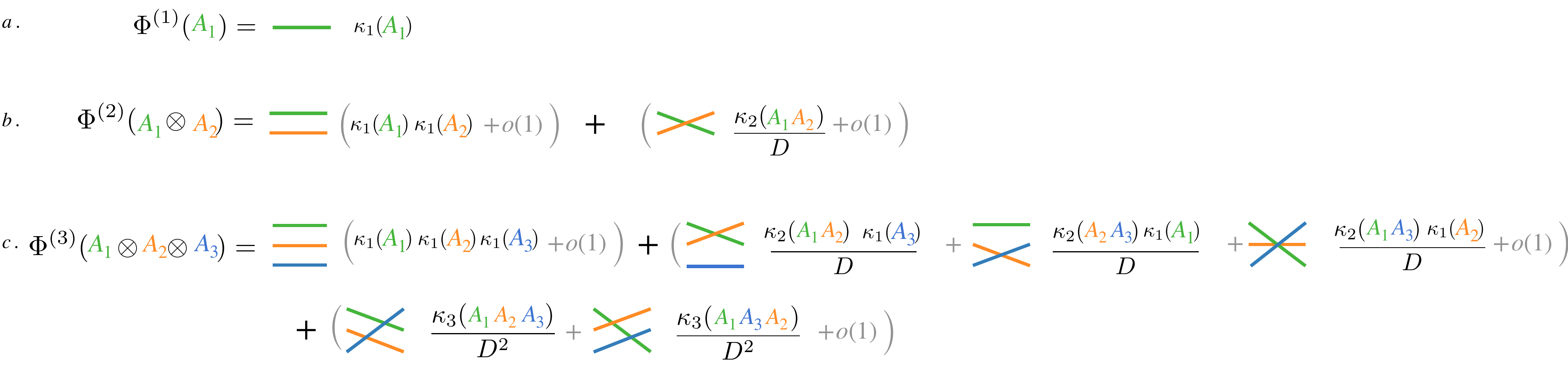}
\end{figure}
\end{widetext}
where the colored lines indicate the permutation operators: for instance, parallel lines correspond to identities, while two lines that cross to the swap $P_{(2, 1)}$ and so on. In the simple case where $A_{i}=A$, these expressions recover the expansions of $\Phi_\text{\rm Haar}^{(k)}$ in Eq.\eqref{eq_PhiHaar} introduced in Sec.\ref{sec_haar}.

Summarizing, it is known that $k$-th Haar channels are the sum of permutations. Here, we have identified coefficients at the leading order in $D$ as free-cumulants. The subleading corrections of order $O(D^{-2})$ are irrelevant for the asymptotic properties of mixed OTOC (as we discuss below).

\subsection{Mixed OTOC of $k$-designs}
We are now in the position to review the derivation of Eq.~\eqref{mainC} \cite{nica2006lectures} using the above rewriting of the $k$-fold quantum channel. 
Multi-point $2k$-mixed OTOC are known to contain information about $k$-channels~\cite{roberts2017chaos}, since they can be re-expressed as
\begin{align}
\label{eq_kOTOC}
 &\langle   A_1^U B_1 \dots A_k^U B_k \, \rangle_{\mathcal E} \nonumber\\
 &\quad= \frac 1D \text{Tr} \left (
	\Phi_{\mathcal E}(A_1 \otimes \dots \otimes A_k)\,  B_1\otimes \dots \otimes B_k\, P_\gamma 
	\right)\ ,
\end{align}
where $P_\gamma$ is the cyclic permutation over $k$ replicas. 

Using the result for the Haar $k$-fold channel~\eqref{eq:mainA_a}, in App.\ref{app_otoc}, we show that $2k$-OTOC can be written as
\begin{align}
\label{eq:k-OTOC-result}
	 &\langle   A_1^U B_1 \dots A_k^U B_k \, \rangle_{\rm Haar} \nonumber\\
	 &\quad= \sum_{\pi \in NC(k)} \kappa_\pi(A_1,\dots,A_k) \, \langle B_1....B_k\rangle_{\pi^*} \ ,
\end{align}	 
which corresponds to the anticipated result of Eq.\eqref{mainC} generalized to different matrices. Thus we recognize that this is the same formula of the product of mixed moments of free variables in Eq.\eqref{eq_free_prod}. The example of an 8-OTOC shown in Fig.\ref{fig:complement} reads
\begin{align}
    \langle A^U_1 B_1&  A^U_2 B_2 A^U_3 B_3 A^U_4 B_4\rangle_{\text{Haar}}   \\
    & = \kappa_1(A_1)\kappa_1(A_2)\kappa_1(A_3) \kappa_1(A_4)\braket{B_1B_2B_3B_3} \nonumber \\
    & +  \kappa_2(A_1,A_2)\kappa_1(A_3)\kappa_1(A_4)\braket{B_1B_2B_3}\braket{B_4} + [3] \nonumber \\
    & + \kappa_2(A_1,A_2)\kappa_2(A_3, A_4)\braket{B_1B_3}\braket{B_2}\braket{B_4} + [1]\nonumber \\
    & + \kappa_3(A_1,A_2,A_3)\kappa_1(A_4)\braket{B_1B_4}\braket{B_2} \braket{B_3} + [3] \nonumber \\
    &    + \kappa_4(A_1,A_2,A_3,A_4) \braket{B_1}\braket{B_2} \braket{B_3}\braket{B_4}
     ,
    \nonumber
\end{align}
where $[n]$ represents the other combination of observables which can be obtained by the same non-crossing partition, i.e. see the $[\times n]$ in Fig.\ref{fig:complement}.

\section{Designs for ETH class of operators}
\label{sec_result_eth}
So far we have introduced the notion of $k$-freeness and showed that for an ensemble $\mathcal{E}$ forming a $k$-design the $2k$-mixed OTOC can be written in terms of free cumulants, yielding an expression identical to when $A^U$ and $B$ are $k$-free. Here we will show that the same applies to unitary evolution with chaotic Hamiltonians at sufficiently long-times. 

\subsection{Time-average ensemble of a Hamiltonian evolution}
In this section, we consider the special ensemble $\mathcal{E}$ of unitaries generated by the time evolution under a fixed chaotic Hamiltonian, i.e., for a fixed Hamiltonian $H$,
\begin{equation}
   \mathcal E_{H, t_{\rm Max}} = \{ U(t)=e^{-iHt} \} \quad \quad \text{with }\quad t \in [0, t_{\rm Max}] \ ,
\end{equation} 
with the time $t$ uniformly distributed in the interval. Hence the average over the ensemble shall be thought of as the standard time-average
\begin{equation}
    \label{time_ave}
     \mathbb E_{t_{\rm Max} }
     (f(U(t))) 
\equiv \frac{1}{t_{\rm Max}} \int_{0}^{t_{\rm Max}}f(U(t)) dt \ .
\end{equation}
Even if $H$ is chaotic, $\mathcal{E}_{H, t_{\rm Max}}$ does not approximate a $k$-design at long times. This was already pointed out in Ref.~\onlinecite{roberts2017chaos}. Formally, this can be seen by explicitly computing the distance induced by the Frobenius norm ($\lVert A \rVert_2 = \sqrt{A^\dag A}$) from the Haar $k$-fold channel $\Phi_{\rm Haar}^{(k)}$ to the channel associated to $\mathcal E_{H, t_{\rm Max} }$.
It turns out that the minimum distance is achieved for large times, at which
\begin{equation}
\label{eq_distance}
    \left\lVert \Phi_{\rm Haar}^{(k)} - \Phi_{\mathcal E_{H,t_{\rm Max} }}^{(k)} \right\rVert_2 \sim \sqrt{k!} D^{k/2}.
\end{equation}

The explicit calculation is reported in Sect 3.3 of Ref.~\cite{roberts2017chaos}.
Intuitively, it is expected that the Hamiltonian evolution is very different from a $k$-design, given that 
the Hamiltonian hosts $D$ conserved quantities, the projectors on the eigenstates. 

Since one deals with Hamiltonian evolution, it is natural to consider the thermal average introduced above, i.e.  
\begin{equation}
\label{eq_thermal}
\langle \bullet \rangle^\beta = \frac{1}{Z} \Tr \left (e^{-\beta H} \bullet \right ) \ ,
\end{equation}
which reduces to the usual definition for $\beta=0$, i.e. $\langle \bullet \rangle^{\beta=0}= \langle \bullet \rangle$.

While the $k$-fold channel acting on a generic set of operators is not close to the Haar $k$-channel [cf. Eq.\eqref{eq_distance}], one could ask if its action on the restricted set of operators $\mathcal A$, namely those satisfying full ETH, is similar to that of the Haar $k$-channel. We will show in the following that this is the case and that Eq.~\eqref{eq:k-OTOC-result} can be recovered from the full ETH formalism of Ref.~\cite{foini2019eigenstate, pappalardi2022eigenstate, pappalardi2023general}.\\

Before discussing long-times, let us remark that these arguments put us in a position to understand easily what becomes of these considerations in the \emph{semiclassical limit}. In fact, all operators of which we normally take semiclassical expectations are by nature ETH-class in $\mathcal A$, they have the requisites of smoothness in the semiclassical limit that guarantee this. What has become then, in the semiclassical limit, of the (vast majority of) operators that are not ETH? A moment's thought tells us that these are operators that have a `bad' semiclassical expression, such as tensor products of two pure states, or 
operators of the form $A^{\frac c \hbar}$.

\subsection{Long-time averages}
Within this spirit, we restrict ourselves to $A, B\in \mathcal A$ and we show below that ETH implies arbitrary $k$-freeness upon time-average, namely
\begin{equation}
\label{eq:long-time-free}
    \mathbb E_{t_{\rm Max}\to \infty} \left [\kappa^\beta_{2k}\left (A(t), B, A(t), B,.... A(t), B \right )\right ] = 0.
\end{equation}
This result holds for \emph{any finite temperature} defining the thermal average in Eq.\eqref{eq_thermal} and for infinite temperature ($\beta=0$) it reduces to the factorization of a $k$-design in Eq.~\eqref{eq:k-OTOC-result}.
 As a consequence of Eq.\eqref{eq:long-time-free}, at infinite times $2k$-OTOCs factorize in terms of moments which depend only on $A$ or $B$ separately, namely:
\begin{equation}
\label{eq:k-OTOC-result-beta}
    \mathbb E_{t_{\rm {Max}\to \infty}}\langle A(t) B ... A(t) B \rangle^\beta = \sum_{\pi\in NC(k)} \kappa^{\beta}_{\pi}(A) \langle B\rangle^\beta_{\pi^*}\ ,
\end{equation}
analogously to the factorization property if $A(t)$ and $B$ were $k$-free, cf. Eq.~\eqref{eq_free_prod}. Note that even if the average over the time-ensemble $\mathbb E_{t_{\rm Max}\to \infty}$ appears in a linear fashion, the thermal free cumulants defined here are sufficient to encode non-linearly the time-evolution ensemble. This is due to the fact that -- at leading order in $1/D$ -- the ``ensemble'' average of the product of moments is equal to the product of averaged moments for $D\to \infty$, as we further elaborate in App.~\ref{sec:app:time-averaging}.

To derive Eq.~\eqref{eq:long-time-free}, we use the result of Ref.~\cite{pappalardi2022eigenstate, pappalardi2023general}, which showed that, when ETH holds, free-cumulants can be rewritten in a simple form in terms of matrix elements of the operators $A$ and $B$ in the Hamiltonian eigenbasis
\begin{align}
    \begin{split}
        \label{eq:k-OTOC-FF-expansion}
    &\kappa_{2k}^{\beta} \left (A(t), B, A(t), B,... A(t), B \right ) \\&= 
    \sum_{\substack{i_1,\dots,i_k\\ \bar{i}_1,\dots,\bar{i}_k\\\text{distinct}}} 
    \frac{e^{-\beta E_{\bar{i}_1}} }{Z}
    A_{\bar{i}_1 i_1} B_{i_1 \bar{i}_2}\dots A_{\bar{i}_{k-1} i_{k}} B_{i_{k} \bar{i}_1} 
    e^{-i t \sum_{l=1}^k (E_{i_l}-E_{\bar{i}_l})} \ .
    \end{split}
\end{align}
Here the indices $i_l$ and $\bar{i}_l$ label the eigenstates of the Hamiltonian and $E_i$ their energies --- we labeled as $\bar{i}$ the eigenstates that correspond to a backward evolution. 
Eq.\eqref{eq:k-OTOC-FF-expansion} shall be considered as a part of the ETH ansatz itself, it has been heuristically argued in Ref.\cite{pappalardi2022eigenstate}.
If we were not restricting the sum to a pairwise distinct set of indices, the above would simply be an expansion of the $2k$-OTOC. The non-trivial aspect is that restricting the sum to pairwise distinct indices isolates the contribution of the $2k$th free cumulant to the OTOC~\cite{pappalardi2023general,pappalardi2022eigenstate}.

When averaging over time in a sufficiently long time window (which we further discuss below), only terms satisfying
\be
    \sum_{l=1}^k E_{i_l} =  \sum_{l=1}^k E_{\bar{i}_l}
\ee
survive. Furthermore, if the system is chaotic, we can assume that there will not be exact resonances in the spectrum and the only way to achieve the equality above is to have the set $\{i_l\}$ be a permutation of the set $\{\bar{i}_l\}$. Therefore, under these assumptions, all the terms in the expansion~\eqref{eq:k-OTOC-FF-expansion} vanish upon time integration.
Concurrently, the factorization~\eqref{eq:k-OTOC-result-beta} applies to the $(2k)$-OTOCs of ETH operators in $\mathcal A$. \\
This is one of the main results of this paper: we have shown that ETH implies that chaotic time-evolution generates arbitrary $k$-free observables for all finite $k$.  

Even if these findings are here derived explicitly for OTOC of only two $A$ and $B$, Eq.\eqref{eq:long-time-free} is expected to \emph{hold also for distinct ETH observables} $A_i, B_i\in \mathcal A$ namely
\begin{equation}
\label{eq:long-time-free_}
    \mathbb E_{t_{\rm Max}\to \infty} \left [\kappa^\beta_{2k}\left (A_1(t), B_1, A_2(t), B_2,.... A_k(t), B_k \right )\right ] = 0 \ .
\end{equation}
In fact, the full ETH in Eq.\eqref{ETHq} holds also for distinct operators \cite{pappalardi2022eigenstate}, implying, therefore, the more general results in Eq.\eqref{eq:long-time-free_}.

\subsection{Physical long-times}

The discussion above tells us that as $t_{\rm Max} \to \infty$, the ensemble $\mathcal{E}_{H, t_{\rm Max}}$ reproduces the Haar ensemble as far as $2k$-OTOCs of $\mathcal A$ ETH-class operators are concerned.
{ Formally, this result is guaranteed to hold when averaging over an exponentially long time-scales.}
Physically, the natural question is whether there is a finite time scale at which the statement above becomes approximately exact not only on average but at each time. We thus define \emph{free}-$k$ time-scale after which alternating free cumulants vanish\footnote{ As mentioned above, since powers of ETH operators are ETH, the condition \eqref{eq:free-k-times} implies that all mixed free cumulants of order smaller than or equal to $2k$ vanish as well.}, namely
\begin{equation}
\label{eq:free-k-times}
   \kappa^\beta_{2k}\left (A(t), B, A(t), B,..., A(t), B \right ) \approx 0\quad \text{for} \quad t\gtrsim t_k.
\end{equation}
Intuitively this is a reasonable expectation in many cases, i.e. far from thermal phase transitions. In ETH the magnitude of matrix elements is assumed to depend smoothly on the energy of the eigenvectors. Therefore, in Fourier space $\kappa^\beta_{2k}(\vec{\omega})$, c.f. Eq.~\eqref{eq:free-cumulant-ETH-intro}, will be a smooth function of the frequency vectors, and the energy scale over which $\kappa^\beta_{2k}(\vec{\omega})$ behaves smoothly will determine $1/t_k$.

The idea that operators in chaotic dynamics are organized in a hierarchy of increasing complexity with $k$ is well established, see e.g. Refs.\cite{roberts2017chaos}. Here, we propose to address this complexity via the value of the $k$th free cumulants.  
The free-$k$ time $t_k$ shall increase with $k$ and may be compared to the design time. 
This picture is consistent with the recent numerical investigation of free cumulants in local many-body systems \cite{pappalardi2023general}, which demonstrated that the $\kappa_4(t)$ is characterized by a much longer time-scale to respect $\kappa_2(t)$, in other words, $t_2\gg t_1$.

The physics is particularly rich in systems in finite spatial dimensions. Here, locality and the finite butterfly velocity introduce extra time scales, that can diverge with the linear system size (i.e. scaling like $\log(D)$), which will modify the smoothness properties of $\kappa^\beta_{2k}$~\cite{chan2019eigenstate}.
Understanding how locality enters in this hierarchy of time scales for generic $k$ remains an interesting challenge.

\subsection{The postulates of Deutsch and Srednicki meet Free Probability}

The basic principles postulated by Deutsch \cite{deutsch1991quantum} and Srednicki \cite{srednicki1994chaos} find a nice and compact expression in terms of Free Probability. 
Consider a generic Hamiltonian  $H$ and perturb it with a non-commuting one $H'$ as
\begin{equation}
    \label{eq_Hlambda}
    H_\lambda = H + c\; \lambda H'\; ;
\end{equation}
 choosing $\lambda$ of $O(1)$, but $c$ as a constant that goes to zero in the thermodynamic limit, e.g., as a power law $N^{-a}$ with $a>0$. The argument underlying ETH is that $H'$ does not change the macroscopic physics of $H$, but is strong enough with respect to the level spacing so as to act as a `blender' for many nearby levels. On the other hand, for finite $\beta$, we may approximate $e^{-\beta H_\lambda}\sim e^{-\beta H}=\rho$ for all $\lambda$ of $O(1)$.

Thus one may consider the unitary $ U_{\lambda}$ which changes the basis between the perturbed and the unperturbed Hamiltonian, namely
\begin{equation}
\label{eq_Ul}
    [\, U_{\lambda}\, ]_{nm} = \bra{E_n}E^{(\lambda)}_m\rangle
\end{equation}
where $E_n$ (and $E^{(\lambda)}_m$) are the eigenvectors of $H$ (and $H_\lambda$) in
Eq.\eqref{eq_Hlambda}. The set of unitaries $ U_{\lambda} $ could now play the role of our \emph{ETH
 ensemble}. 
 Differently from fully random unitaries, this matrix is structured in energy, with matrix elements that decay at large $\omega=E_n-E_m$, as depicted on the right-hand side of Fig.\ref{fig_summary}.
Thus, rotational invariance (that characterizes Haar rotations) emerges here only close to the diagonal -- for small frequencies -- leading to the concept of \emph{local rotational invariance}, which has been used to deduce the energy-dependent ETH correlations in Ref.\cite{foini2019eigenstate}. See also Refs.\cite{pappalardi2022eigenstate, shi2023local} for other discussions on $U_\lambda$.

Even if the unitaries $U_{\lambda} $ are clearly not a group, they enjoy another important property. If we 
consider a set $\{H_{\lambda_1},...,H_{\lambda_k}\}$ with different perturbations $\lambda_i$, the corresponding observables $A$ written in the basis of macroscopically indistinguishable Hamiltonians:
$\{A_{\lambda_1} ,...,A_{\lambda_k}\}$
are mutually free, with measure $\rho$. In other words, different elements of the $U_\lambda$ ensemble \eqref{eq_Ul} generate mutually free observables. 

Understanding the impact of mutual freeness and other properties of the ETH ensemble on designs is an interesting question which is left as a direction for future work.

{
\section{Outlook: thermalization for ensembles of states}
\label{sec_ensStates}
We now discuss how the formalism introduced before may be used to address questions related to thermalization, such as if
time-evolved states
\be
|\psi(t)\rangle = e^{-i H t} |\psi(0)\rangle \
\ee
can be considered as ``pseudo-random''.

 To give a precise meaning to the term pseudorandom, we consider \emph{ensemble of states}
 $   \mathcal S_{|\psi\rangle}  = \{ p_j, |\psi_i\rangle \}\ ,$
where $|\psi_i\rangle$ are random states distributed with probability $p_i$.  The information of the full distribution is encoded in the moments
\begin{equation}
    \label{rho_k}
    \rho_\mathcal S^{(k)} = \mathbb E_{\psi\sim \mathcal S}[(|\psi\rangle\langle \psi|)^{\otimes k}] = \sum_i p_i (|\psi_i\rangle\langle \psi_i|)^{\otimes k}\ .
\end{equation}
An ensemble $\mathcal S$ is called a \emph{quantum state design} if its $k$-th moment \eqref{rho_k} equals the $k$-th moment of a 
random ensemble, i.e.
$ \rho_\mathcal S^{(k)}= \rho_{\rm Haar}^{(k)}$.
The rationale for the name $\rho_{\rm Haar}^{(k)}$ derives from the fact that
random states can always be written as the application of random unitaries $U_i$ to a reference state $\ket{0}$, i.e. $|\psi_i\rangle = U_i |0\rangle$. It is then clear that information about the ensemble of states $\mathcal{S}=\{p_i, U_i|0\rangle\}$ is included in the knowledge about the ensemble of unitaries $\mathcal{E}=\{p_i, U_i\}$ that we discussed above. For instance, Eq~\eqref{rho_k} can be written in terms of the $k$-fold quantum channel as
\begin{equation}
    \label{eq_RE_PhiE}
    \rho_{\rm Haar}^{(k)} = \Phi^{(k)}_{\rm Haar}(|0\rangle\langle 0|^{\otimes k})\ .
\end{equation}

{As mentioned above, one would like to understand to what extent unitary evolution generated by a chaotic Hamiltonian can lead to a quantum state design. For this purpose, we define the ensemble of states
\ba
\mathcal{S}_{H, t_{\rm Max}} &= \big\{\ket{\psi(t)} =  e^{-i H t} |\psi(0)\rangle \big\}
&
\text{with }t&\in[0,t_{\rm Max}] 
\end{align}
and ask whether at long times it shares properties of a quantum state design.
Clearly $\mathcal{S}_{H, t_{\rm Max}}$ cannot form a quantum-state design, not even for $k=1$ since the time evolution conserves many quantities, such as the projectors onto all its eigenstates, hence the energy.

  In this outlook, we consider some simple instances (such as moments or R\'enyi entropies) to illustrate how free probability can be used to show that \emph{quantum state designs are asymptotically equivalent to time-evolved states at long times}.
For an exhaustive discussion on the topic, we refer to Refs.\cite{kaneko2020characterizing, pilatowsky2023complete, 2024arXiv240311970M}.
Finally, we conclude by discussing possible connections and differences between $k$-freeness and the recently-introduced notion of ``deep thermalization''.\\
}
}

{
\subsection{ Moments of expectation values}

 We begin discussing moments of observable expectation values, whose information is encoded in the $k$-th moment in Eq.\eqref{rho_k}, i.e.
\begin{align}
    \label{fluctua-mom}
    \mathbb E_{\psi\sim \mathcal S} & \left[
\bra{\psi} A\ket{\psi} ^k
    \right ] 
    = \Tr \left ( \rho_\mathcal S^{(k)} A^{\otimes k}\right ) \ .
\end{align}
In the case of a quantum-state design, we can now use the free-probability approach outlined in Sec.~\ref{sec_result_des}.
A straightforward calculation (see App.\ref{app_fluactua}) at the leading order in $1/D$ leads to
\begin{align}
    \label{fluctua}
    \mathbb E_{\psi\sim \rm Haar} & \left[ \bra{\psi} A\ket{\psi} ^k
    \right ] 
    = \Tr \left ( |0\rangle\langle 0|^{\otimes k} \Phi^{(k)}_{\rm Haar} (A^{\otimes k})  \right ) \nonumber
    \\ &
    \simeq \kappa_1(A)\dots \kappa_1(A) + \mathcal O(D^{-1}) \ .
\end{align}
Hence, the average of the $k$-th power of an expectation value is dominated by the $k$-th power of the average of the observable. Since the fluctuations are subleading at large $D$, they are not accounted for by our asymptotic approach.

The same happens for chaotic dynamics on $\ket{\psi(t)}$: the time evolution at infinite times leads to asymptotic state designs \emph{on the diagonal ensemble identified by the initial state $\ket{\psi(0)}$},  i.e.
\begin{equation}
    \label{diag}
    \langle \bullet \rangle_{\rm diag} = \Tr(\rho_{\rm diag} \bullet) \quad \text{with}\quad \rho_{\rm diag} = \sum_{i} |c_i|^2 \ket{E_i} \bra{E_i}
\end{equation}
where $ c_i = \langle \psi(0) | E_i\rangle$ are the overlaps with the initial state, which, for generic initial states, can be assumed to be well peaked around an extensive energy $E_0$, with sub-extensive energy fluctuations. 

The $k$-th moment of observables on the time-dependent state can be re-written as
 \begin{align}
        \label{eq:k-mom-expan}
    &\langle \psi(t) |A|\psi(t)\rangle^k\\&= 
    \sum_{\substack{i_1,\dots,i_k\\ \bar{i}_1,\dots,\bar{i}_k}} 
    c_{i_1} c^*_{\bar i_1}\dots c_{i_k} c^*_{\bar i_k}
    A_{i_1 \bar{i}_1 } \dots A_{i_{k}\bar{i}_{k} }
    e^{-i t \sum_{l=1}^k (E_{\bar i_l}-E_{{i}_l})} \ .
\nonumber
\end{align}
Averaging over infinite times and using the non-resonance condition discussed in Sec.\ref{sec_result_eth}, in App.\ref{app_momETH} we show that ETH implies
 \begin{align}
        \label{eq:k-mom-expansion}
    &\mathbb E_{t_{\rm Max} \to \infty}[\langle \psi(t) |A|\psi(t)\rangle^k]
     \simeq \langle A\rangle_{\rm diag}^k + O(e^{-S}) \ .
\end{align}

{Summarizing, we have discussed how the
$k$-th moments of physical observables are asymptotically the same for quantum state designs and time-evolved states at large times:  they are given by the $k$-th power of the first cumulant, i.e.  $[\kappa_1(A)]^k = \langle A \rangle^k$, the latter being evaluated over the diagonal ensemble for Hamiltonian time-evolution.} Note that Eq.\eqref{fluctua} and Eq.\eqref{eq:k-mom-expansion} are immediately generalized to the fluctuations of the product of expectation values of different observables, such as $\langle A^{(1)}\rangle \dots \braket{A^{(k)}}$.

It is worth noting that for different energy distributions, one might compare the time-average ensemble with more sophisticated choices of the random ensemble, see e.g. Ref.~\onlinecite{2024arXiv240311970M}.

\subsection{R\'enyi entropies}

We will now discuss the R\'enyi entropies across a bipartition between two complementary regions $X$ and $\bar X$. The total Hilbert space is factorized $\mathcal H=\mathcal H_X \otimes \mathcal H_{\bar X}$ and has dimensions $\dim \mathcal H_X = D_X$ and $ \dim \mathcal H_{\bar X}=D_{\bar X}= D/D_X$. Given a pure state $\ket{\psi} \in \mathcal H=\mathcal H_X \otimes \mathcal H_{\bar X}$ and 
its the reduced density matrix $\rho_X= \Tr_{\bar X} |\Psi\rangle\langle \Psi|$,  we are interested in the higher order purity
\begin{equation}
    \label{Renyi}
    R^{(k)}(|\Psi\rangle) = \Tr(\rho_X^k) \ .
\end{equation}
Using the partial cycle permutation $S_X$, defined as the operator that cyclically permutes the $k$ replicas over the subsystem $X$, i.e. 
    $S_X |i_1, \bar i_1; i_2, \bar i_2; \dots i_k \bar i_k\rangle = 
    |i_2, \bar i_1; i_3, \bar i_2; \dots i_1 \bar i_k\rangle $ \footnote{Here $|i_{1, \dots k}\rangle$ ($|\bar i_{1, \dots k}\rangle$) represents the basis of $\mathcal{H}_X$ ( $\mathcal{H}_{\bar X}$).}, Eq.\eqref{Renyi} can be re-written as
    \begin{equation}
    \label{Renyi_}
    R^{(k)}(|\Psi\rangle) = \langle \Psi|^{\otimes k} S_X |\Psi\rangle^{\otimes k}\ ,
\end{equation}
see e.g. Ref.\cite{kaneko2020characterizing}. The $k$-th R\'enyi entropy are the defined from Eq.\eqref{Renyi} as $S^{k} = \frac 1{1-k} 
 \log R^{(k)}$\footnote{Rather than being concerned with the average R\'enyi entropies of an ensemble of states, we will consider the average of their exponentials, which directly probe the $k$-th moment of the state, c.f. Eq.~\eqref{rho_k}.}. 

Free Probability, despite being asymptotic, allows to show that \emph{quantum state designs and a time-average ensemble possess the same R\'enyi entropy at the leading order}, including the so-called Page correction, also known as ``residual entropy''~\cite{page1993average}. \\
The R\'enyi entropy of random Haar states has been intensively studied, see e.g.~\cite{zyczkowski2001induced, kim2024average, collins2010random, collins2011gaussianization, liu2018entanglement}, including via Free Probability tools \cite{collins2010random, collins2011gaussianization, liu2018entanglement}. In App.\ref{app_Renyi}, we provide an alternative derivation using asymptotic free probability. The $k$-th purity in Eq.\eqref{Renyi} reads
\begin{align}
\label{eq:renyi_fp_result}
    \mathbb E_{\psi\sim \rm Haar}[ \Tr(\rho^{k}_X)]    
    & = \sum_{\pi \in NC(k)} \frac 1{D_{X}^{k-\#\pi^*}}\, \frac 1{D_{\bar X}^{k-\#\pi}} + \mathcal O(D^{-1})\ ,
\end{align}
where $\#\pi$ and $\#\pi^*$ are the number of blocks of the partition $\pi$ and of its dual.
This expression is symmetric upon exchanging $X$ and $\bar{X}$. Moreover, when $X$ and $\bar X$ become comparable, i.e. $D_X=D_{\bar X}=\sqrt D$, it gives 
\begin{align}
\label{eq:renyi_result_catalan}
    \mathbb E_{\psi\sim \rm Haar}[ \Tr(\rho^{k}_X)]    
    & = \frac {C_k}{ D_{X}^{k-1}}+ \mathcal O(D^{-1})\ ,
\end{align}
where $C_k$ is the Catalan number which counts the non-crossing partitions of $NC(k)$. Upon taking the logarithm, this contribution leads to the so-called { residual entropy} of the R\'enyi entropies: $S^{(k)} = \ln D_{X} - \frac{\ln C_k}{k-1} + \mathcal O((\ln D)^{-1})$ \cite{collins2010random, collins2011gaussianization, liu2018entanglement}. 

 Let us comment now on how one may apply these findings to the
time-averaged entropy of the non-equilibrium state, defined as
\be
    \overline{R^{(k)}} = \frac{1}{t_{\rm Max}} \int_0^{t_{\rm Max}} \dd t \, R^{(k)}(\ket{\psi(t)})\ .
\ee
Using Eq.\eqref{Renyi_}, we compute the long-time average for large enough $t_{\rm Max}$, by expanding in the energy eigenbasis. With the assumption of the absence of exact resonances, c.f. Sec.~\ref{sec_result_eth}, the average over $\exp(-i t (E_{j_1} - E_{\bar j_1} +\dots+E_{j_k} - E_{\bar j_k}))$ selects only permutations of the energy indices. This leads to 
\ba
    \mathbb E_{t_{\rm Max} \to \infty} &[R^{(k)}(|\psi(t)\rangle )] = 
    \sum_{i_1, \dots i_k} |c_{i_1}|^2 \dots |c_{i_k}|^2 \,
    \\
     & \quad \times \sum_{\beta \in S_k} \langle E_{i_1} \dots E_{i_k}| P_\beta S_X |E_{i_1} \dots E_{i_k}\rangle \ ,
\nonumber
\end{align}
where $c_i=\langle E_i|\psi(0)\rangle$ is the diagonal ensemble in Eq.\eqref{diag}. 
To make further progress,
one may consider the case where the coefficients $|c_j|^2=1/D$ for all $j$.
Here, one has 
\ba
\label{eq:renyi-eth-flat}
\overline{R^{(k)}}&=D^{-k}\sum_{\beta \in S_k} \Tr(P_\beta S_X)
\end{align}
(cf. Eq.\eqref{eq:renyi_haar_result_} of App.\ref{app_Renyi}), thus recovering the result~\eqref{eq:renyi_fp_result} for a quantum state design.

 This result is expected to apply 
to generic states, where $|c_j|^2\sim 1/D$ averaged over small energy windows, also known as infinite-temperature initial states.
In the spirit of ETH, we expect $\langle E_{i_1} \dots E_{i_k}| P_\beta S_X |E_{i_1} \dots E_{i_k}\rangle$ to be independent of small variations in the set of indices $i_1,\dots,i_k$, but rather to only have a smooth dependence on the energies $E_{i_1}$, \dots, $E_{i_k}$. In this way, small-scale fluctuations of $|c_j|^2$ do not modify the result in Eq.~\eqref{eq:renyi-eth-flat}.

More generally, it would be interesting to better understand how this result generalizes to finite-temperature states. There, we expect energy conservation to affect the average value of $R^{(k)}$. In fact, it is known that the entropy of finite-energy state has additional corrections due to the curvature of the density of states, c.f. Refs.~\onlinecite{PhysRevE.100.022131,PhysRevLett.119.220603}.


\subsection{$k$-freeness vs. ``deep thermalization''}
\label{sec_deep}
The notion of $k$-freeness introduced in this paper is one way to extend the usual concept of thermalization. Indeed, the thermalization of the local density matrix 
corresponds to the case $k=1$ discussed in this manuscript.
This is however not the only possible generalization. For example, recent advances in quantum simulators stimulated the idea of ``deep thermalization''~\cite{cotler2023emergent, Choi_2023, ho2022exact, claeys2022emergent, ippoliti2022solvable,ippoliti2023dynamical,lucas2022generalized, bhore2023deep, mcginley2022shadow, PRXQuantum.4.030322}.

For concreteness consider a lattice system, which is bipartite into two regions $S$ and $R$. The regular notion of thermalization involves the reduced density matrix on $S$ obtained after tracing over $R$, i.e. $\rho_S = \Tr_R \ket{\psi(t)} \bra{\psi(t)}$. To define deep thermalization, rather than simply tracing over $R$, we perform projective measurements onto each site of $R$. In this way, the state over $R$ is collapsed onto a product state, specified by the array of measurement outcomes $\vec{m}$. In general, we then have a random state of the form $\ket{\phi_{t,\vec{m}}}_S \otimes \ket{\vec{m}}_R$ selected with probability $P_{\vec{m}} = \braket{\psi(t)|\vec{m}}_R {}_R\braket{\vec{m}|\psi(t)}$. The ensemble of states $\ket{\phi_{t,\vec{m}}}_S$ and their probability $P_{\vec{m}}$ forms the so-called \emph{projected ensemble}. Deep thermalization concerns whether this ensemble forms a quantum state design. 

$k$-freeness and deep thermalization clearly involve different quantities and concepts. Nonetheless one might ask if there is a common physical origin. While this is an important direction for
future work, we here discuss an example based on the results of Ref.~\onlinecite{2023arXiv230508437S} that hints towards a \emph{negative answer: While $k$-freeness has the same properties of unitary dynamics, such as locality, deep thermalization is intrinsically non-local.}

One way to address this question quantitatively is by studying if there is a relation between deep-thermalization times $\tilde{t}_k$ and free-$k$ times $t_k$. 
 Firstly,
one must introduce a notion of locality within $k$-freeness. In the bipartition setting above this can be most easily done by redefining $t_k$ as the maximum time overall set of possible operators $A$ and $B$ with support in $S$ to achieve the condition~\eqref{eq:free-k-times}. In the specific setting of Ref.~\onlinecite{2023arXiv230508437S} where $S$ sits between two $R$ regions, we consider how $t_k$ and $\tilde{t}_k$, obtained for open boundary conditions (BCs), can be modified upon adding a coupling between the first and final $R$ sites.
In this case, changing the BCs cannot affect $t_k$. The Lieb-Robinson bound~\cite{lieb1972finite} implies that $A(t)$ has support over a finite region and thus is independent of the BCs in the thermodynamic limit. Therefore, if $t_k$ is finite in the thermodynamic limit, it must be BC-independent.
On the other hand, this is not the case for deep thermalization at $k\geq 2$. Even in the thermodynamic limit and at finite times the projected ensemble can drastically depend on the BCs. A remarkable example is the one presented in Ref.~\onlinecite{2023arXiv230508437S}, where deep thermalization takes place at finite times, but nonetheless using periodic BCs halves $\tilde{t}_k$ w.r.t. the open BCs case.
The origin of the non-locality is the combination of the measurements and the sharing (classical communication) of the measurement outcomes over the whole system. 

While this specific example indicates that in general there will not be a connection between $k$-freeness and deep thermalization, it would nonetheless be interesting to understand if there are limiting cases where these two phenomena share the same mechanism and timescales.
}

\section{Conclusions and perspectives}
\label{sec_conclusion}
In this paper, we propose to consider designs from the point of view of Free Probability. In this language, an operator $A$ and $U A U^\dagger$, the latter rotated with an element of a $k$-design,  are $k$-free. This implies that the $2k$-OTOCs factorize and the mixed free cumulants vanish.

A by-product of our work is the new notion of $k$-freeness, which, in turn, is accompanied by all the tools of Free Probability now available in connection with $k$ designs. \\

{
The results of this work have broader consequences for the chaotic dynamics of many-body systems:
\begin{itemize}
    \item ETH equipped with Free Probability allows one to identify the \emph{universality of late-time chaotic dynamics in the factorization of higher moments} into products of free cumulants,  each of them with observables evaluated at the same time. This acts as a sort of generalization of the mixing condition \cite{lebowitz1973modern}. We have shown that unitary evolution always generates $k$ free variables at $k$ sufficiently long separated times. In other words, the notion of freeness gives a compact and operational way to understand the effect of chaotic time evolution by vanishing the appropriate free cumulants of all orders.
    \item While other approaches to this issue are based on frame potentials or operator's norms, this point of view shifts the focus to \emph{moments and cumulants of physical observables}. This must be considered as being in the original spirit of ETH, which takes us from postulating an ensemble of Hamiltonians having the same macroscopic physics, yet are random `in in the small' (Berry and Deutsch), to the ensuing thermalization properties. This is a natural step, as when one goes from a classical description in terms of trajectories in phase space to one in terms of the evolution of physical observables.
    \item This approach provides \emph{practical tools} to verify when time-evolution yields a $k$-design, studying at what times the free cumulants vanish:
    \[
    \kappa_{2k}(A, A(t), \dots  A, A(t) ) = 0\ .
    \]
    This condition is easily implementable on classical simulations of quantum dynamics and may be experimentally achievable via some of the techniques to measure OTOC, see e.g. \cite{li_measuring_2017, garttner_measuring_2017, landsman_verified_2019, joshi_quantum_2020, blok_quantum_2021}. 
    \item The definition of $k$-freeness does not rely on the existence of eigenstates and it {naturally applies to arbitrarily time-dependent and noisy evolutions} (such as random circuits). As such, the vanishing-free cumulants above may be easily computed in a much larger class of quantum evolutions whose chaotic properties we may thus probe.
    \item Furthermore, this line of thought naturally includes the ETH ideas on \emph{equilibration at the level of ensembles of states}. Our free probability approach shows that the chaotic time evolution of pure states (equipped with the appropriate diagonal ensemble) is equivalent to quantum state designs from the point of view of observable fluctuations or entanglement entropies.  
\end{itemize}    
}

{ As mentioned at the outset, all the new notions introduced in this work may, and perhaps sometimes need, be relaxed by replacing equalities with equalities up to a predefined precision $\epsilon$. Our approach is based on asymptotic definitions, however a rigorous definition of an $\epsilon$-approximated $k$-freeness would require a thorough discussion of the finite $D$ corrections. While in the case of full-matrices, further insights in this direction could be gained using the so-called `second-order freeness' \cite{collins2024second}, in the ETH case it remains an open question what to use as an effective dimension. 
Finite-size corrections may be relevant also in the discussion of fluctuations of multi-time observables such as $\mathbb E[\langle A(t) A\rangle^k]$.

Physically, 
an important question concerns the implications of locality on the meaning and scaling of the free-$k$ times $t_k$. In this direction, it would be very interesting to understand what are the implications of $k$-freeness on the study of transport. \\
This problem is linked to the relationship between $k$-freeness and deep-thermalization. As discussed in Section \ref{sec_deep}, these two are generally distinct issues, but it would be worthwhile to investigate if there are limiting cases where the two phenomena share mechanisms or time scales.\\
Finally, an interesting but challenging question concerns the fate of late-time $k$-freeness when ETH does not apply, such as for different forms of ergodicity breaking or in the case of integrable systems \cite{essler2023statistics}.

}

\section*{Acknowledgements}
We thank Laura Foini for insightful comments and collaboration on related works. SP thanks Xhek Turkeshi for the initial insights on the Haar ensemble and for useful discussions.  We thank Takato Yoshimura for valuable feedback on the first version of Sec.~\ref{sec_result_des}.
We would also like to thank Andrea de Luca, Jacopo de Nardis, Denis Bernard, and Wen Wei Ho for valuable discussions on the topic.
This work was completed during the workshop ``Open QMBP'' held at the Institut Pascal at Université Paris-Saclay made possible with the support of the program “Investissements d’avenir” ANR-11-IDEX-0003-01.  S.P. acknowledges support by the Deutsche Forschungsgemeinschaft (DFG, German Research Foundation) under Germany’s Excellence Strategy - Cluster of Excellence Matter and Light for Quantum Computing (ML4Q) EXC 2004/1 -390534769.

\medskip
\appendix

\section{Examples of Haar channels} 
\label{app:sec}
 The Haar quantum channel can be written explicitly by using the Schur-Weyl duality \footnote{The Schur-Weyl duality states that an operator $A$ on $\mathcal H^{\otimes k}$ commutes with all operators $V^{\otimes k}$ with $V\in \mathcal U(\mathcal H)$ if and only if it is a linear superposition of permutation operators}:
\begin{equation}
	\Phi_{\rm Haar}^{(k)}(O) = \sum_{\alpha \beta \in S_k} \text{Wg}_{\alpha, \beta}(D)\,\text{Tr}(P_\beta O) \, P_{\alpha^{-1}}\ ,
\end{equation}
where $\text{Wg}_{\alpha, \beta}(D)$ is the Weingarten matrix, defined as the inverse of the Gram matrix \cite{collins2003moments, gu2013moments}\footnote{{We assume the existence of the inverse $Q^{-1}$, valid for $k\leq N$, which is the case under consideration.}}
\begin{equation}
	Q_{\alpha, \beta} =\text{Tr}( P_\alpha P_\beta) = D^{\#(\alpha^{-1}\beta)} \ ,
\end{equation}
where $\#(\alpha)$ counts the number of cycles of the permutation $\alpha$. Let us show explicitly the first few $k$: \\

\noindent \textbf{$\bullet {\bf k=1}$}: $S_1 =\{(1)\}$, $P_{\alpha^{-1}} = \{\mathbb I\}$ $Q_{11}=d$, $\text{Wg}_{11}(D) = \frac 1D$ and 
\begin{equation}
\Phi_{\rm Haar}^{(1)}(O) = \frac {\text{Tr}(O)} D \mathbb I = \langle A \rangle \mathbb I\ ,
\end{equation}
where we defined the normalized expecation, for $A\in \mathcal H$
\begin{equation}
\langle A \rangle =  \frac {\text{Tr}(A)} D	 \ .
\end{equation}

\noindent \textbf{$\bullet {\bf k=2}$}: $S_2 = \{(12), (21)\}$, $P_{\alpha^{-1}}=\{P_{12}, P_{21}  \}$
\[
Q = \begin{pmatrix}
	D^2 & D\\
	D & D^2
\end{pmatrix}
\quad \quad 
\text{Wg}= \frac 1{D(D^2-1)}\begin{pmatrix}
	D & - 1 \\
	- 1  & D
\end{pmatrix}\ ,
\]
which leads to
\begin{align}
\begin{split}
\Phi_{\rm Haar}^{(2)}(O) & = \frac {D \text{Tr}(O) - \text{Tr}(O\, P_{12})}{D(D^2-1)} P_{12}\\ & \quad 
+  \frac {D \text{Tr}(O \, P_{12}) - \text{Tr}(O)}{D(D^2-1)}\, P_{21} \ .    
\end{split}
\end{align}
If $O=A\otimes A$, then we have 
\[
 \text{Tr}(O) =  \text{Tr}(A^{\otimes 2})= \text{Tr}(A)^2 = D^2 \langle A\rangle^2\ ,
 \]
 \[
  \text{Tr}(O\, P_{21}) =  \text{Tr}(A^{\otimes 2} \, P_{21})= \text{Tr}(A^2) = D \langle A^2\rangle\ ,
\]
which leads to
\begin{align}
    \begin{split}
\Phi_{\rm Haar}^{(2)}(A^{\otimes 2})&  = \frac {D^2 \langle A\rangle^2 - \langle A^2\rangle}{(D^2-1)} P_{12}
+  \frac { D \langle A^2\rangle - D \langle A\rangle^2 }{(D^2-1)} \, P_{21}
\\ & =_{D\to \infty}
\langle A\rangle ^2\, P_{12} + \frac{ \langle A^2\rangle -  \langle A\rangle^2}D \, P_{21}\ .
    \end{split}
\end{align}

\noindent \textbf{$\bullet {\bf k=3}$}: the permutation group is $S_3 = \{(123), (213), (321), (132), (231), (312)\}$,
\begin{equation}
Q = \begin{pmatrix}
	D^3 & D^2 & D^2 & D^2 & D & D\\
	D^2 & D^3 & D & D & D^2 & D^2\\
	D^2 & D & D^3 & D & D^2 & D^2\\
	D^2 & D & D & D^2 & D^2 & D^2\\
	D & D^2 & D^2 & D^2 & D^3 & D\\
	D & D^2 & D^2 & D^2 & D & D^3\\
\end{pmatrix}	
\end{equation}
and
\begin{align}
\text{Wg} & = \frac 1{(D^2-1)(D^2-4)} \times \nonumber \\
&  \times
\begin{pmatrix}
	1 - \frac 2D & -1 & -1 & -1 & \frac 2D & \frac 2D\\
	-1 &1 - \frac 2D & \frac 2D & \frac 2D& -1 & -1 \\
	-1 &\frac 2D &1 - \frac 2D &  \frac 2D& -1 & -1 \\	
	-1 &\frac 2D &\frac 2D& 1 - \frac 2D &  -1 & -1 \\	
	\frac 2D &  -1 & -1 & -1& 1 - \frac 2D  &\frac 2D \\	
	\frac 2D &  -1 & -1 & -1 &\frac 2D& 1 - \frac 2D 
\end{pmatrix}\ ,	
\end{align}
Which for large $D$ results in 
\begin{align}
    \begin{split}
        \label{Phi3}
	\Phi_{\rm Haar}^{(3)}(A^{\otimes k}) & =_{D\to \infty} \langle A\rangle^3 P_{123}
\\ &\quad +
\langle A \rangle ( \langle A^2\rangle -  \langle A\rangle^2) \frac{ P_{213}+ \text{perm.}}D 
\\ &\quad +
(\langle A^3\rangle -3 \langle A\rangle \langle A^2\rangle + 2\langle A\rangle^3  ) \frac{ P_{231}+ P_{312}}{D^2} 
    \end{split}
\end{align}

It is clear that the new information contained in the $k$-th channel that was not in the $k-1$ is associated with the cyclic permutation $P_{12\dots k}$ and has the form of a connected correlation function. To understand this structure we will need Free Probability.

\section{Free Probability on $k$-designs}

\subsection{Derivation of Eq.~\eqref{mainA}}
\label{app_deri}
Let us here illustrate the main steps leading to the expansion of the $k$-fold Haar quantum channel in terms of free cumulants. 
\paragraph{Scaling with $D$.---}The starting point to show Eq.~\eqref{mainA} is to expand the Weingarten matrix in Eq.~\eqref{eqWg} to leading order in $D$ \cite{collins2003moments}
\begin{equation}
    \label{Wg_D}
	\text{Wg}_{\alpha, \beta}(D)\simeq \frac{1} { D^{2k-\#(\beta^{-1}\alpha)}} \left[ \mu(\beta, \alpha)  { + O(D^{-2})}\right] \ .
\end{equation}
To extract the leading contributions, it is necessary to make the $D$-dependence explicit, i.e.
\be
\label{eq_traceWb}
\Tr (P_\beta A_1\otimes A_2\otimes \dots \otimes A_k) = D^{\# \beta} \prod_{b\in \text{Orb}(\beta)} \langle A_{j_1}\dots A_{j_b}\rangle.
\ee
In the expression above $b$ runs over the orbits of $\beta$ -- in other words, the cycles of $\beta$. It is further intended that $j_1, \dots, j_b$ enumerate the element of the orbit in the appropriate order, i.e.
$A_{j_2}= A_{\beta(j_1)}$, $A_{j_3}= A_{\beta(j_2)}$, and so on, until $A_{j_{1}}= A_{\beta(j_{|b|})}$.

The two equations lead to
\begin{align}
    \Phi_{\rm Haar}^{(k)}(A_1\otimes \dots A_k) & = \sum_{\alpha  \in S_k} P_{\alpha^{-1}} \sum_{\beta \in S_k} \frac{\mu(\beta, \alpha)}{ D^{2k-\#(\beta^{-1}\alpha) - \#\beta}} 
    \nonumber \\ & 
    \quad \times  \prod_{b\in \text{Orb}(\beta)} \langle A_{j_1}\dots A_{j_b}\rangle + O(D^{-2})
    \label{eq_sum}
\end{align}
Therefore, focusing on a given $P_{\alpha^{-1}}$ its coefficient $\kappa_\alpha$ at leading order is determined by the permutations $\beta$ that minimize $2k-\#(\beta^{-1} \alpha)-\#\beta$. 
To understand which permutations $\beta$ give the leading contributions,
we recall the notion of \emph{length $\ell(\sigma)$ of a permutation} $\sigma$, i.e. the least number of two-element transpositions necessary to factorize $\sigma$ \cite{nica2006lectures}. 
 This is useful for our purposes since the length is directly related to the number of cycles through $\ell(\sigma)= k - \# \sigma$.
 Furthermore, the length can be used to define a distance on $S_k$
$
    d(\alpha,\beta) = \ell(\alpha^{-1} \beta).
$
In particular, we will use that $d$ satisfies the triangular inequality. In our case, considering the distance between $\alpha$ and the identical permutation $\id$, we have that $\ell$ satisfies
\be
\label{eq_triag}
    \ell(\beta) + \ell(\beta^{-1} \alpha) \geq \ell(\alpha)
\ee
From this, it follows that the permutations $\beta$ giving the leading contributions are the ones saturating the triangular inequality, i.e.
\begin{equation}
    \label{eq_sat_traing}
    \ell(\beta) + \ell(\beta^{-1}\alpha) = \ell(\alpha)\ ,
\end{equation}
whereas all the other permutations can be neglected.
With reference to the distance $d$ induced by $\ell$, the set of permutations $\beta$ satisfying~\eqref{eq_sat_traing} is often called \emph{geodesic}; in this case the geodesic between $\alpha$ and the identity $\id$.
Eq.~\eqref{eq_sat_traing} constrains the sum over $\beta$ in Eq.\eqref{eq_sum} and leads to a coefficient proportional to $1/D^{\ell(\alpha)}=1/D^{k-
\#\alpha}$ with corrections $O(D^{-2})$. Putting everything together, we then obtain Eqs.~\eqref{mainA} with
\begin{multline}
	\label{mainB2}
	\kappa_\alpha(A_1,A_2,\dots ,A_k) = \sum_{\beta: \ell(\beta) + \ell(\beta^{-1}\alpha) = \ell(\alpha)}
	\mu(\beta, \alpha)\\
    \times \prod_{b\in \text{Orb}(\beta)} \langle A_{j_1}\dots A_{j_b}\rangle .
\end{multline}
\\

\paragraph{The coefficients $\kappa_\alpha$ and non-crossing partitions}

The next question we want to ask is whether there is a more explicit way of writing $\kappa_\alpha$. In fact, we will now show that $\kappa_\alpha$ is simply related to the free cumulants $\kappa_\pi$ previously introduced.

This connection hinges on an important combinatorial property of the distance $d(\alpha, \beta) = \ell(\alpha^{-1}\beta)$, leading to
an \emph{embedding of non-crossing partitions into the permutation group}.
We begin by restricting ourselves to the case where the permutation $\alpha$ satisfies the conditions (i) and (ii) listed above Eq.~\eqref{eq:kappa-alpha-kappa-pi}. Namely, the cycles of $\alpha$ induce a non-crossing partition $\pi$ and permute the elements of each cycle counterclockwise w.r.t. our drawing convention. Then, the permutations $\beta$  contributing to $\kappa_\alpha$, i.e. those lying on the geodesic between the identity $\id$ and $\alpha$, 
can be characterized as the set of permutations $\beta$ such that: the orbits of $\beta$ induce a non-crossing partition $\sigma$, $\sigma\leq\pi$ (each block of $\sigma$ is contained in a block of $\pi$), and all cycles of $\beta$ permute their elements counterclockwise w.r.t. our convention.
For further details and a proof, we refer to the Lecture 23 of Ref.\cite{nica2006lectures}.
This immediately leads to the result of Eq.\eqref{mainA} and to identification of $\kappa_\pi(A_1, A_2, \dots A_k)$ with the free cumulants.\\

The previous statement can be extended to permutations which have a different cyclic order. 
This can be done by noting through Eq.~\eqref{eq:permutation-conjugation}, which we also report below for the reader's convenience
\be
    \kappa_{\alpha}(A_1\cdots A_k) = \kappa_{\rho^{-1} \alpha \rho} (A_{\rho(1)}\cdots A_{\rho(k)}),\, \forall \rho \in S_k.
\ee
To show this, we observe that
\be
    \Phi^{(k)}_{\rm Haar}(A_1 \otimes \cdots \otimes A_k) = P_{\rho}^\dag  \Phi^{(k)}_{\rm Haar}(A_{\rho(1)} \otimes \cdots \otimes A_{\rho(k)}) P_{\rho}.
\ee
Expanding both sides using Eq.~\eqref{mainA}, then yields Eq.~\eqref{eq:permutation-conjugation}. Eq.~\eqref{eq:permutation-conjugation} is particularly useful since for any $\alpha$, we can always choose $\rho$ such that $\alpha'=\rho^{-1} \alpha \rho$ has non-crossing cycles that permute their elements counterclockwise (see e.g. Fig.~\ref{fig:permutations}(c) for an example). [In fact two permutations, $\alpha$ and $\alpha'$, can always be related by conjugation $\rho^{-1} \alpha \rho$ if the lengths of their cycles are the same.]
Hence, with an appropriate choice of $\alpha'$
we can reduce ourselves to the case studied in the previous paragraph.
Correspondingly, the coefficient $\kappa_\alpha$ in Eq.\eqref{mainA} is given by the free cumulant of the associated permutation:
\begin{align}
	\kappa_\alpha(A_1,A_2,\dots ,A_k)& = \sum_{\sigma \leq \pi} 
	\mu(\pi, \sigma)  \langle A_{\rho(1)},A_{\rho(2)},\dots ,A_{\rho(k)}\rangle_\sigma \nonumber \\
 & = \kappa_\pi(A_{\rho(1)},A_{\rho(2)},\dots ,A_{\rho(k)}) \
  .
\end{align}

\subsection{Derivation of Eq.\eqref{eq:k-OTOC-result}}
\label{app_otoc}
We now want to show how $2k$-OTOC can be written in the form~\eqref{mainC} using the result for the Haar $k$-fold channel~\eqref{eq:mainA_a}.
For this purpose, we start from the $2k$-OTOC in Eq.~\eqref{eq_kOTOC} and we plug in the result of the Haar $k$-th channel of Eqs.~\eqref{eq:mainA_a} and~\eqref{mainB2} as
\begin{widetext}
\begin{subequations}
\begin{align}
	  \langle   A_1^U B_1 \dots A_k^U B_k \, \rangle_{\rm Haar} 
	& = \frac 1D \text{Tr} \left (
	\Phi_{\rm Haar}(A_1 \otimes \dots \otimes A_k)\,  B_1\otimes \dots \otimes B_k P_\gamma 
	\right)
 \\ &
	= \frac 1D \sum_\alpha \kappa_\alpha(A_1, \dots ,A_k) \frac{\text{Tr} \left ( P_\gamma P_{\alpha^{-1}}  B_1\otimes \dots \otimes B_k\right ) }{D^{k-\#\alpha}}
	\\
	& = \sum_\alpha \frac{\kappa_\alpha(A_1, \dots ,A_k)}{D^{k+1-\#\alpha -\#(\gamma\alpha^{-1})}} \, \prod_{b \in \text{Orb}(\gamma \alpha^{-1})} \langle B_{j_1}\dots B_{j_b}\rangle \ ,
 \label{eq_54c}
\end{align}    
\end{subequations}
\end{widetext}
where from the first to the second line we have used Eq.\eqref{mainA} and then we made the $D$ dependence explicit using Eq.\eqref{eq_traceWb} for $\gamma \alpha^{-1}$. For large $D$, the leading contribution to this summation is given by the permutations $\alpha$ which minimize $k+1-\#\alpha -\#(\gamma\alpha^{-1})$. Using that the length of the cyclic permutation is $\ell(\gamma) = k-1$, it follows that the leading contribution satisfy:
\begin{equation}
    \label{eq_la}
    \ell(\alpha)+ \ell(\gamma\alpha^{-1}) = \ell(\gamma)\ .
\end{equation}
Which implies that in the large $D$ limit, Eq.\eqref{eq_54c} can be re-written as
\begin{align}
\label{eq:k-OTOC-permuta}
	 &\langle   A_1^U B_1 \dots A_k^U B_k \, \rangle_{\rm Haar} \\
	 &\quad= \sum_{\alpha: \ell(\alpha)+ \ell(\gamma\alpha^{-1}) = \ell(\gamma)} \kappa_\alpha(A_1, \dots, A_k) \langle B_1....B_k\rangle_{\gamma \alpha^{-1}}\nonumber
\end{align}	
where we used the definition of $\langle B_1...B_k\rangle_{\gamma \alpha^{-1}}$ as the products of moments one for each block of the permutation, e.g.  given $\langle B_1...B_k\rangle_{\gamma \alpha^{-1}}=
\prod_{d\in\text{Orb}(\gamma \alpha^{-1})}\langle \prod_{j\in d} B_j\rangle
$.  

The condition \eqref{eq_la} implies that $\alpha$ lies on the geodesic between the identity and the cyclic permutation $\gamma$. Hence, due to the embedding of the non-crossing partitions discussed above, $\text{Orb}(\alpha)=\pi$ is a non-crossing partition. Furthermore, one can show that $\gamma \pi^{-1}= \pi^{*}$ corresponds to the dual of non-crossing partition of $\pi$ \cite{nica2006lectures}.
Note that, from this discussion, it is clear that all permutations $\alpha$ contributing to OTOC satisfy the conditions (i) and (ii) discussed above in Sec.\ref{sec_haarkfold}. Hence, together with Eq.\eqref{eq:kappa-alpha-kappa-pi}, this leads to Eq.\eqref{eq:k-OTOC-result}.

\section{Time-average of the cumulant vs cumulant of the time-averaged ensemble}
\label{sec:app:time-averaging}

In Sec.~\ref{sec_result_eth} we showed that the average of the free cumulant defined according to Eq.~\eqref{freekETH} is zero in the long-time limit, c.f. Eq.~\eqref{eq:long-time-free}. 
However, the time average of the cumulant $\kappa^\beta_{2k}$ is distinct, in principle, from the cumulant of the time average ensemble $\mathcal{E}_{H, t_{\rm Max}}$, which in this appendix we will simply denote as $\mathcal{E}$. For the purpose of this appendix, we will explicitly denote the ensemble in the latter with $\horror_{2k}$. Furthermore, to show that $\mathcal{E}_{H, t_{\rm Max}}$ acts as a $k$-design on ETH-class operators, we need to show that $\horror_{2k}$ vanishes in the long-time limit.
In this appendix, we demonstrate that, if the average of $\kappa^\beta_{2k}$ is zero~\eqref{eq:long-time-free}, then also $\horror_{2k}$ is zero, up to corrections of size $1/D$.

We first begin by showing that $\horror_{2k}$ and the time average of $\kappa^\beta_{2k}$ do not coincide in general. The simplest example of this is obtained for $k=2$. Writing explicitly $\horror_{2k}$ using Eq.~\eqref{eq_k4} we have
\begin{align}
    \horror_4(& A(t) B A(t) B) \nonumber \\
    & =
    \mathbb{E}_{t_{\rm Max}}\left[\langle A(t)BA(t)B \rangle^\beta\right] \nonumber\\
    &~~ - 2 {\color{red} \left(\mathbb{E}_{t_{\rm Max}}\left[\langle A(t) B \rangle^\beta \right]\right)^2 }
    \nonumber\\
    &~~ - 2 \mathbb{E}_{t_{\rm Max}}\left[ \langle A(t) B A(t) \rangle^\beta \right] \langle B \rangle^\beta \nonumber\\
    &~~ - 2 \mathbb{E}_{t_{\rm Max}}\left[ \langle B A(t) B \rangle^\beta \right] \langle A \rangle^\beta
    \nonumber\\
    &~~ + 10 \mathbb{E}_{t_{\rm Max}}\left[ \langle A(t) B \rangle^\beta \right]  \langle A \rangle^\beta \langle B \rangle^\beta \nonumber\\
    &~~ - 5 \left(\langle A \rangle^\beta\right)^2 \left(\langle B \rangle^\beta\right)^2
\end{align}
These are the same terms that contribute to $\mathbb{E}_{t_{\rm Max}}\left[ \kappa^\beta_4(A(t) B A(t) B) \right]$, except that the term highlighted in red above would be replaced by
\begin{equation}
    \mathbb{E}_{t_{\rm Max}}\left[ \left( \langle A(t) B \rangle^\beta \right)^2 \right] \neq \left(\mathbb{E}_{t_{\rm Max}}\left[\langle A(t) B \rangle^\beta \right]\right)^2.
\end{equation}

Nonetheless, at long times, and at leading order in $1/D$ we will argue that the time average can be split as
\begin{align}
\label{eq:app:approximation-general}
    \mathbb{E}_{t_{\rm Max}} & \left[ \langle A(t) B \cdots \rangle^\beta \cdots \langle A(t) B \cdots \rangle^\beta \right] 
    \\ = & 
    \mathbb{E}_{t_{\rm Max}}\left[ \langle A(t) B \cdots \rangle^\beta\right] \cdots 
    \mathbb{E}_{t_{\rm Max}}\left[\langle A(t) B \cdots \rangle^\beta \right] + O(1/D),\nonumber
\end{align}
where each expectation value might contain an arbitrary number of terms.

For concreteness, we will show this explicitly in the simplest possible case, i.e.
\begin{multline}
\label{eq:app:approximation}
    \mathbb{E}_{t_{\rm Max}}\left[ \langle A(t) B \rangle^\beta \langle A(t) B \rangle^\beta \right] \\
    =\mathbb{E}_{t_{\rm Max}}\left[ \langle A(t) B \rangle^\beta\right] 
    \mathbb{E}_{t_{\rm Max}}\left[\langle A(t) B \rangle^\beta \right] 
    \\ + O(1/D),
\end{multline}
but the argument extends straightforwardly to the most general case.
We expand the LHS in the eigenbasis of the Hamiltonian as
\begin{multline}
\label{eq:app:eigenstate-expansion}
    \mathbb{E}_{t_{\rm Max}} \left[ \langle A(t) B \rangle^\beta \langle A(t) B \rangle^\beta \right]     \\  
    = 
    \sum_{\substack{i,j\\ \bar{i},\bar{j}}} 
    \frac{e^{-\beta (E_{\bar{i}} + E_{\bar{j}})} }{Z^2}
    A_{\bar{i} i} B_{i \bar{i}}  A_{\bar{j} j} B_{j \bar{j}} \\
    \times \mathbb{E}_{t_{\rm Max}}\left[ e^{-i t (E_{i}-E_{\bar{i}} + E_{j}-E_{\bar{j}})} \right]\ .
\end{multline}
Proceeding as in Sec.~\ref{sec_result_eth}, at long times, assuming that there are not exact resonances in the spectrum,
\begin{multline}
\label{eq:app:average}
    \mathbb{E}_{t_{\rm Max}}  \left[ e^{-i t (E_{i}-E_{\bar{i}} + E_{j}-E_{\bar{j}})} \right] \\
    =
    \delta_{i,\bar{i}} \delta_{j,\bar{j}} + \left( \delta_{i,\bar{j}} \delta_{j,\bar{i}} - \delta_{i,\bar{i}} \delta_{j,\bar{j}} \delta_{i,j} \right)\,.
\end{multline}
Plugging the first of the two terms in Eq.~\eqref{eq:app:eigenstate-expansion} we obtain exactly the factorized term
$\mathbb{E}_{t_{\rm Max}}\left[ \langle A(t) B \rangle^\beta\right] \mathbb{E}_{t_{\rm Max}}\left[\langle A(t) B \rangle^\beta \right]$.
Instead, plugging the second term in Eq.~\eqref{eq:app:eigenstate-expansion} we obtain a ``crossing'' partition contribution, namely:
\begin{equation}
    \sum_{\substack{i\neq\bar{i}}} 
    \frac{e^{-\beta (E_{\bar{i}} + E_{i})}}{Z^2}
    \left( A_{\bar{i} i} B_{i \bar{i}} \right)  \left(A_{i \bar{i}} B_{\bar{i} i} \right) \sim 1/D \ ,
\end{equation}
which is suppressed as $1/D$ \cite{pappalardi2022eigenstate}, as can be seen using the standard ETH scaling  $\overline{A_{\bar{i} i} B_{i \bar{i}}} \sim e^{-S(E_+)}$,
with $E_+=(E_i+E_{\bar{i}})/2$. 
Therefore the approximation in Eq.~\eqref{eq:app:approximation} holds.

The same arguments work in general at all orders. Expanding the LHS of Eq.~\eqref{eq:app:approximation-general} in terms of eigenstates, there will be the average of oscillating phases as in Eq.~\eqref{eq:app:average}. The average of oscillating phases will always yield pairing schemes where the energy levels are equal within a given expectation value, but otherwise unconstrained, and terms constraining levels from different expectation values coincide. The former terms give the factorized contribution in the RHS of Eq.~\eqref{eq:app:approximation-general}. The latter is instead always to be suppressed by factors of $1/D$ or higher.

Finally, let us remark that the ``asymptotic'' factorization between ensemble averages of products and products of ensemble averages [cf. Eq.\eqref{eq:app:approximation}] holds in the same way for the Haar ensemble. 

{
\section{Free Probability on ensembles of states}
In this appendix, we report the main steps in the application of Free Probability theory to ensembles of states discussed in Section \ref{sec_ensStates}.

\subsection{Calculation of Eq.\eqref{fluctua}}
\label{app_fluactua}
We derive the leading contribution to the moments of expectation values using Free Probability. 
First of all, we rewrite Eqs.~\eqref{eq_RE_PhiE} and~\eqref{fluctua-mom} using the cyclicity of the trace and invariance of the Haar distribution under $U\mapsto U^\dag$. In this way, we obtain
\footnote{Alternatively, one might have tried to apply the result of Sec.~\ref{sec_result_des} by starting with the state in Eq.~\eqref{eq_RE_PhiE} and applying the formula for the Haar channel in Eqs.\eqref{mainB}-\eqref{eq:mainA_a} with $A_1=\cdots=A_k=\ket{0}\bra{0}$.
Note, however, that the approach in Sec.~\ref{sec_result_des} was developed for operators $A_j$ s.t.
$\Tr(A_j)\sim O(D)$, hence those results cannot be applied directly in this case, and it is necessary to proceed as described in the main text.}

\begin{align}
    \label{fluctua_}
    \mathbb E_{\psi\sim \rm Haar} & \left[ \bra{\psi} A\ket{\psi} ^k
    \right ] 
    = \Tr \left ( |0\rangle\langle 0|^{\otimes k} \Phi^{(k)}_{\rm Haar} (A^{\otimes k})  \right ) \nonumber
    \\ & \simeq  \sum_{\alpha \in S_k} \frac{\kappa_\alpha(A, \dots, A)}{D^{k-\#\alpha}} \Tr( P_{\alpha^{-1}}  |0\rangle\langle 0|^{\otimes k} )  + \mathcal O(D^{-2}) \nonumber
    \\ & =   \sum_{\alpha \in S_k} \frac{\kappa_\alpha(A, \dots, A)}{D^{k-\#\alpha}}  + \mathcal O(D^{-2})
    \\ &
    \simeq \kappa_1(A)\dots \kappa_1(A) + \mathcal O(D^{-1}) \nonumber\ ,
\end{align}
with $\kappa_1(A)=\Tr(A)/D$. Here, from the first to the second line we used Eq.\eqref{eq:mainA_a} and to the third $\Tr(P_\alpha  |0\rangle\langle 0|^{\otimes k}) =1 \quad \forall \alpha$. Finally we note that the leading contribution comes from the identity permutation $P_\alpha=\mathbb I$, while the other terms are suppressed as $D^{-1}$.

\subsection{Derivation of Eq.\eqref{eq:k-mom-expan}}
\label{app_momETH}
We here derive the infinite-time limit of the $k$-th moment of observables in Eq.\eqref{eq:k-mom-expan}.
As discussed  in Sec.\ref{sec_result_eth}, when averaging over infinite times, i.e. using Eq.\eqref{time_ave},
only terms satisfying   $\sum_{l=1}^k E_{i_l} =  \sum_{l=1}^k E_{\bar{i}_l}$
survive and, using the non-resonance condition, this equality above is to have the set $\{i_l\}$ be a permutation of the set $\{\bar{i}_l\}$. This leads to 
 \begin{align}
        \label{eq:k-mom-expansion_}
    &\mathbb E_{t_{\rm Max} \to \infty}[\langle \psi(t) |A|\psi(t)\rangle^k]
    \\&= \label{all_indices}
    \sum_{\substack{i_1,\dots,i_k\\ \alpha \in S_k}} 
    |c_{i_1}|^2\dots |c_{i_k}|^2 
    A_{i_1 \alpha({i}_1) } \dots A_{i_{k}\alpha({i}_{k})} 
    \\ & \simeq \langle A\rangle_{\rm diag}^k + O(e^{-S}) \ . \nonumber
\end{align}
To obtain the last line above, one recognizes that the leading contribution to Eq.\eqref{all_indices} is given by the identity permutation:
\begin{equation}
    \left (\sum_{i_1} |c_{i_1}|^2 A_{i_1i_1}\right )^k = \langle A\rangle_{\rm diag}^k\ ,
\end{equation}
while, for $\alpha$ different from the identity, one can use ETH to bound the other terms as
\begin{align}
    \sum_{\substack{i_1,,i_k \rm dist.\\ \alpha \in S_k}} &
    |c_{i_1}|^2\dots |c_{i_k}|^2 
    A_{i_1 \alpha({i}_1) } \dots A_{i_{k}\alpha({i}_{k})} \nonumber
    \\ & 
    \leq \max A_{i_1 \alpha({i}_1) } \dots A_{i_{k}\alpha({i}_{k})} 
    \\ &
    \propto e^{-\#(\text{OD})S}
\end{align}
 where $\#(\text{OD})$ counts the number of off-diagonal matrix elements included in the products, viz. $\#(\text{OD})=\#\{j|\alpha(j)\neq j\}$. Here $S$ is the minimum of the thermodynamic entropy where $|c_j|$ are non-negligible.  

\subsection{R\'enyi entropy for state designs using Free Probability}
\label{app_Renyi}
We here provide a derivation of Eq.\eqref{eq:renyi_fp_result} of the main text, see also Refs.\cite{collins2010random, collins2011gaussianization, liu2018entanglement}. The average of the purity $R^{(k)}$ over a state design \eqref{eq_RE_PhiE} reads
\begin{align}
    \label{renuirandom}
    &\mathbb E_{\ket{\psi}\sim \rm Haar}[ \Tr(\rho^{k}_X)]   
    =  \Tr \left ( |0\rangle\langle 0|^{\otimes k} \Phi^{(k)}_{\rm Haar}(S_X) \right)   
    \\ & = 
    \sum_{\alpha \beta \in  S_k} {\rm Wg}(\alpha, \beta^{-1}) \Tr(P_{\alpha^{-1}}|0\rangle\langle 0|^{\otimes k})\, \Tr(P_\beta S_X) \ .
    \nonumber
\end{align}
In the second line, we have used the exact result for the quantum channel in Eq.\eqref{eqWg}.  We recall that the starting point of our Free Probability approach was the asymptotic expansion for the Weingarten matrix in Eq.\eqref{Wg_D}
\begin{align}
    {\rm Wg} (\alpha, \beta^{-1})& \simeq \frac{\mu(\alpha \beta^{-1})}{D^{2k-\#\alpha \beta^{-1}}}
    \simeq \frac{\delta_{\alpha \beta}}{D^k} + \mathcal O(D^{-(k+1)}) \ ,
    \label{Wg_DD}
\end{align}
where in the second line we have used that the leading contribution to this matrix is diagonal in the replicas. While for observables the off-diagonal terms lead to finite results, we will now show that the diagonal approximation is sufficient to get the Page correction to the entanglement \cite{page1993average}. 
We plug Eq.\eqref{Wg_DD} into Eq.\eqref{renuirandom}, together with $\Tr(P_{\alpha^{-1}}|0\rangle\langle 0|^{\otimes k})=1$ $\forall \alpha$, this leads to
\begin{align}
\label{eq:renyi_haar_result_}
    \mathbb E_{\psi\sim \rm Haar}[ \Tr(\rho^{k}_X)]    
    & \simeq \frac 1{D^k}\sum_{\beta \in S_k}\Tr(P_\beta S_X)
    \\ & 
    =     \sum_{\beta \in S_k} \frac{D_X^{\# \gamma\beta^{-1}} D_{\bar X}^{\#\beta}}{D^k}
\end{align}
where from the first to the second line we used $\Tr(P_\beta S_X)=D_{X}^{\#\gamma \beta^{-1}}  D_{\bar X}^{\# \beta}$, with $\gamma$ is the cycle permutation over the $k$ replicas with the same ciclicity as $S_X$. We now substitute in the denominator $D = D_X D_{\bar X}$ and recognize that the exponents of $D_{X/ \bar X}$ are given by the permutation length function $\ell(\sigma) = k - \#\sigma$ (introduced above in App.\ref{app_deri}). This leads to
\begin{align}
    \mathbb E_{\psi\sim \rm Haar}[ \Tr(\rho^{k}_X)]    
    & 
    \simeq   \sum_{\beta \in S_k} \frac 1{D_X^{\ell(\gamma \beta^{-1})}}
    \frac 1{D_{\bar X}^{\ell (\beta)}}\ .
\end{align}
This expression is maximal when $D_X=D_{\bar X}=\sqrt D$. Hence, 
in the large $D$ limit, the leading contribution to the summation is given by permutations 
that saturate the triangular inequality for the length function in Eq.\eqref{eq_triag}, which in this case reads
\[
\ell(\beta) + \ell(\gamma \beta^{-1}) = \ell(\gamma)\ .
\]
As discussed above, this implies that the permutations $\beta$ lie on the geodesic between the identity and the cyclic permutation. Thus, they correspond to non-crossing partitions and one arrives to the desired result 
\begin{align}
\label{eq:renyi_haar_result__}
    \mathbb E_{\psi\sim \rm Haar}[ \Tr(\rho^{k}_X)]    
    & 
    \simeq   \sum_{\pi \in NC(k)} \frac 1{D_X^{k - \#\pi^*}}
    \frac 1{D_{\bar X}^{k-\#\pi}}\ .
\end{align}
For instance, at $k=2,3,4$  Eq.\eqref{eq:renyi_haar_result__} reads
\begin{subequations}
\begin{align}
     \mathbb E_{\psi\sim \rm Haar}[ \Tr(\rho^{2}_X)]  & 
     =  \frac{1}{D_X} + \frac{1}{D_{\bar X}} \\
     \mathbb E_{\psi\sim \rm Haar}[ \Tr(\rho^{3}_X)]  & = \frac{1}{D_X^2} + \frac{3}{D_XD_{\bar X}} + \frac{1}{D_{\bar X}^2}\\
    \mathbb E_{\psi\sim \rm Haar}[ \Tr(\rho^{4}_X)]  & = \frac{1}{D_X^3} + \frac{6}{D^2_XD_{\bar X}} + \frac{6}{D_XD_{\bar X}^2} + \frac{1}{D_{\bar X}^3}\ .
\end{align}    
\end{subequations}
 As expected, these expressions are symmetric upon exchanging $X$ and $\bar{X}$.  Moreover, when $X$ and $\bar X$ are the same ($D_X=D_{\bar X}=\sqrt D$), it is well-known that the R\'enyi entropies of random states read
   $$S^{(k)}= \frac 1{1-k}\ln     \mathbb E_{\psi\sim \rm Haar}[ \Tr(\rho^{k}_X)] = \ln D_X + c^{(k)}_{\rm Page} + O((\ln D)^{-1})\ ,
   $$
where $c^{(k)}_{\rm Page}$ is a number of order one, known as the Page correction or residual entropy, see e.g. Refs.\cite{page1993average, zyczkowski2001induced, kim2024average, collins2010random, collins2011gaussianization, liu2018entanglement, collins2010random, collins2011gaussianization, liu2018entanglement}.
In this case Eq.\eqref{eq:renyi_haar_result__} can be re-written as
 \begin{align}
\label{eq:renyi_haar_resultPage}
    \mathbb E_{\psi\sim \rm Haar}[ \Tr(\rho^{k}_X)]    
    & 
    \simeq   \frac 1{\sqrt{D}^{k-1}}\sum_{\pi \in NC(k)}  =  \frac {C_k}{\sqrt{D}^{k-1}}
\ ,
\end{align}
where one uses that $\#\pi + \#\pi^* = k+1$ and $C_k$ denotes the number of non-crossing partitions of $NC(k)$, known as the Catalan number. 
In the first few orders, this reads
 \begin{subequations}
\begin{align}
     \mathbb E_{\psi\sim \rm Haar}[ \Tr(\rho^{2}_X)]  & 
     =  \frac{2}{\sqrt D} \\
     \mathbb E_{\psi\sim \rm Haar}[ \Tr(\rho^{3}_X)]  & = \frac{5}{\sqrt D^2} \\
    \mathbb E_{\psi\sim \rm Haar}[ \Tr(\rho^{4}_X)]  & = \frac{14}{\sqrt D^3} \ .
\end{align}    
\end{subequations}
 Finally, Eq.\eqref{eq:renyi_haar_resultPage}  allows us to identify the Page correction:
 \begin{equation}
     c^{(k)}_{\rm Page} = \frac {\ln C_k}{1-k}\ .
 \end{equation}
}

\bibliography{refs}

\end{document}